\documentclass{aa}  
\usepackage[usenames,dvipsnames]{xcolor}

\usepackage{graphicx}
\usepackage{txfonts}
\usepackage{multirow}
\usepackage{pbox}

\usepackage{xspace}
\usepackage{multicol}
\usepackage{siunitx}
\usepackage{natbib,twoopt}
\usepackage{hyperref} 

\usepackage{etoolbox}

\newcommand{\meightonefour}{${m}_{\textrm{F814W}}\thinspace$}
\newcommand{\threethreesix}{${\textrm{F336W}}\thinspace$}
\newcommand{\eightonefour}{${\textrm{F814W}}\thinspace$}

\newcommand{\Vcritfrac}{${v_{\textrm{rot}}}/{v_{\textrm{crit}}}\thinspace$}

\newcommand{\Msun}{\ensuremath{\,\mathrm{M_\odot}}\xspace}

\newcommand{\Msol}{\ensuremath{\,\mathrm{M_\odot}}\xspace}

\newcommand{\GG}[1]{}

 \def\simle{\mathrel{\hbox{\rlap{\hbox{\lower4pt\hbox{$\sim$}}}\hbox{$<$}}}}
 \def\simgr{\mathrel{\hbox{\rlap{\hbox{\lower4pt\hbox{$\sim$}}}\hbox{$>$}}}}

\graphicspath{%
  {./figs/}%
  }

\makeatletter

\begin{document}

   \title{A stringent upper limit on Be star fractions produced by binary interaction}

   \author{B. Hastings  \inst{1,2}
		\and
            N. Langer \inst{1,2}
		\and
            C. Wang \inst{1}
            \and
            A. Schootemeijer \inst{1}
			\and 
			A. P. Milone \inst{3}
            }
    \institute{Argelander-Institut f\"{u}r Astronomie, Universit\"{a}t Bonn, Auf dem H\"{u}gel 71, 53121 Bonn, Germany\\ 
         \and     
    Max-Planck-Institut   f\"{u}r   Radioastronomie,   Auf   dem   H\"{u}gel   69, 53121 Bonn, Germany         \\
           \and 
           Dipartimento di Fisica e Astronomia “Galileo Galilei” —Univ. di Padova, Vicolo dell’Osservatorio 3, Padova, I-35122, Italy 
             }
            
    \authorrunning{Hastings et al.}         

 
\abstract
   {Binary evolution can result in fast-rotating stars, predicted to be observable as Be stars, through accretion of angular momentum during mass-transfer phases. Despite numerous observational evidence pointing to this possibly being the dominant Be formation channel, current models struggle to produce a satisfactory description of Be star populations.}
   {Given distinct uncertainties in detailed binary evolution calculations, we investigate a rigorous and model independent upper limit for the production of Be stars through binary interaction and aim to confront this limit with observations of Be stars in young star clusters. }
   {Using extreme assumptions, we calculate the number ratio of post-interaction to pre-interaction binary systems in a coeval population, which describes an upper limit to Be star formation through mass-transfer. A detailed comparison is made between our derived upper limit and relevant observations of Be stars, which allows us to probe several aspects of binary star physics.  }
   {We find that in coeval populations, binary interaction can at most account for one third of all {main-sequence} stars being Be stars. Near the cluster turn-off region, this limit appears to be realised in the clusters studied. Away from the turn-off, applying simple assumptions about which systems undergo unstable mass-transfer produces a good fit to the observed Be fraction as a function of mass. }
{ We find that assuming distinct physics, binary evolution alone can in principle match the high numbers of Be stars observed in open clusters. Whether the required binary physics is realised in nature remains to be investigated.}


   \maketitle
%

\section{Introduction \label{sec:introduction}}

Be stars are massive main-sequence stars which display emission features in their spectra. While, since their discovery over 150 years ago \citep{BeDiscovery}, we have advanced our understanding to explain the emission as a result of a decretion-disc which is being ionised by the central star \citep{Struve}, it is still not clear how a Be star gains its disc. Observations conclusively show that Be stars rotate significantly faster than their B counterparts \citep{Struve,1996MNRAS.280L..31P,HuangGiesObs,2016A&A...595A.132Z}, such that potentially, the centripetal force matches the gravitational force at the equator \citep{1995ApJ...439..860C,Townsend}. However the fundamental origin of this fast rotation is still unknown, although single and binary star channels have been proposed. 


One way to achieve such rotation is for a star to be spun up by mass-transfer in a binary system \citep{1975BAICz..26...65K,PolsBinaryModels,2006A&A...455.1165L,2012ARA&A..50..107L}. When a star accretes material, it also accretes angular momentum, which in the absence of tidal forces can lead to critical rotation of the accreting star, allowing material to become unbound and form a disc. The accretion of angular momentum is an efficient process, with a star needing to accrete typically a few percent of its own mass to rotate critically \citep{1981A&A...102...17P}.

In wide systems which initiate mass-transfer after the primary has exhausted hydrogen in the core (so-called Case B mass-transfer), tidal forces are generally weak for the accretor star and it can be spun up to near-critical velocities. Furthermore, rapid rotators can also originate from close systems which undergo mass-transfer while the donor is still core hydrogen burning (so-called Case A mass-transfer). Although tides inhibit the spin-up of the accretor during the initial mass-transfer phases, these phases cause a widening of the binary \citep{2005A&A...435.1013P} so that when the donor expands to become a giant star (initiating Case AB mass-transfer), many systems are wide enough to render tides ineffective, allowing the mass-gainer to rotate super-synchronously (Sen et al., in prep.). Therefore rapidly rotating mass-gainers can originate from both short and long period systems. What is common between these cases is that the spun-up star is usually produced after the initially more massive star in the system has exhausted its supply of core hydrogen.

In Be star producing binary systems where the primary star is not massive enough to undergo a supernova explosion, a short-lived helium star or long-lived white-dwarf would be the companion to the Be star. Despite the difficulty of detection, both of these types of systems have been observed \citep{2012ApJ...761...99L,2018A&A...615A..30S,2020A&A...639L...6S,2020MNRAS.497L..50C}. Furthermore, studies of Be star discs have found that many are truncated, suggesting that they are acted upon by unseen companions \citep{2017A&A...601A..74K,2019ApJ...885..147K}.

When the mass-donor does explode as a supernova, the majority of systems are expected to become unbound \citep{1995MNRAS.274..461B} and the Be star will probably have no companion. The fact that this does not occur in every case is evidenced by large numbers of Be-Xray binaries \citep{BeXRBcat}, which consist of a neutron star in an eccentric orbit around a Be star such that Xrays are produced when the Be disc and neutron star interact. When the binary is disrupted, the Be star would likely be a runaway star. \citet{2018MNRAS.477.5261B} and \citet{2020arXiv200903571D} both find the peculiar space-velocities of Be stars in the Gaia catalogue to be consistent with a binary origin of Be stars.  

Observations show the Be phenomenon to be more common at lower metallicities  \citep{MaederObs,MatayanObs,IqbalObs}, in good agreement with predictions of single star models whereby metal-rich stars suffer stronger angular momentum losses through winds, thus making fast-rotators rarer at higher metallicities \citep{2020A&A...633A.165H}. Naively, this trend is difficult to explain in the binary framework. However, further observational characteristics of Be stars have been uncovered that are difficult to explain with a single star formation channel. Initially fast rotating single stars are expected to exhibit enhanced surface nitrogen abundances, as rotational mixing dredges up CNO processed material to the photosphere. However, there appears to be an incompatibility between models of rotating single stars and measurements of nitrogen abundances in Be stars, with many Be stars showing much lower nitrogen abundances than expected \citep{2005A&A...438..265L,2011A&A...536A..65D,2017MNRAS.471.3398A,2020A&A...633A.165H}. On the other hand, spun-up mass-gainers might not be rich in surface nitrogen. Although the physics governing the details of mass-transfer remains uncertain, accretion may be limited by the angular momentum content of the gainer, such that accretion becomes non-conservative once critical rotation is achieved \citep{2020ApJ...888L..12W,2020A&A...638A..39L}. Another factor is the strong mean molecular weight barrier established from hydrogen burning which prevents efficient rotational mixing in the critically rotating mass-gainer \citep{1974IAUS...66...20K,1989ApJ...338..424P}.

 As demonstrated by \citet{2008A&A...478..467E} and \citet{2020A&A...633A.165H}, single stars may achieve near-critical rotation during the late stages of hydrogen burning, in contrast to observations showing that Be stars have a range of fractional main-sequence ages \citep{2005A&A...441..235Z,2005ApJS..161..118M,MiloneObs}. If Be stars are mostly single, one should expect  pre-interaction binaries to host Be primaries, as whatever proposed single star mechanism causes the Be phenomenon should work for stars in a pre-interaction binary just as well as for single stars. It is thus telling that almost no Be stars with a main-sequence companion have been detected \citep{2020arXiv200613229B}. 
 

Despite the numerous pieces of evidence to support the dominance of a binary formation channel, several uncertainties in binary evolution prevent a solid and accurate theoretical description of Be star populations. Proof of the difficulty in modelling the production of Be stars is given by the contrasting results of previous authors. It has been concluded that binaries are responsible for either all \citep{2014ApJ...796...37S}, half  \citep{PolsBinaryModels} or only a small minority \citep{1997A&A...322..116V} of galactic Be stars. This difference is mostly due to different assumptions on mass-transfer efficiency and the stability of mass-transfer.  

In light of these uncertainties, we find it useful to determine a model free upper-limit to Be star production from mass-transfer in binary systems. Assessment of this limit can provide insight into whether it is at all possible for Be stars to be formed exclusively in binaries, and to what extent other formation mechanisms must be invoked. Under the assumption that binary evolution dominates the production of Be stars, we can also probe uncertain binary physics. We shall use recent high quality observations of Be stars in open clusters \citep{MiloneObs} to give a stringent test to our simple picture.

In Section \ref{sec:method} we explain our procedure for calculating an upper-limit to Be star production from mass-transfer in binary systems, with the results of this endeavour presented in Section \ref{sec:results}. In Section 
\ref{sec:obscomp} we compare our results to the numbers of Be stars observed in young open clusters. We infer the conditions for stable mass-transfer that are required for our prescription to reproduce the Be fractions along the main-sequences of young open clusters in Section \ref{sec:stable_conditions}. Uncertainties and the implications of the upper-limit are discussed in Section \ref{sec:disc}. Concluding remarks are given in Section \ref{sec:conc}. 




%

\section{Method \label{sec:method}}
\subsection{A hypothetical population of interacting binary stars}
In order to calculate an upper-limit to the numbers of Be stars that may be produced, we take extreme assumptions. The first of which is that the initial binary fraction in the population is 1; that is every star is born as a member of a binary. Nextly, as the hydrogen-burning episode of a massive star makes up around 90\% of the star's total lifetime, we shall assume that as soon as a primary star leaves the main-sequence, stable mass-transfer will occur on a very short timescale, instantly producing a Be star. In our model, a Be star shall be produced regardless of the initial period, primary mass, or mass-ratio of the system, so that every secondary star will at some point during its lifetime become a Be star. In this framework, the orbital period distribution becomes irrelevant. Furthermore we shall assume that once a Be star is formed, it remains so for the rest of its lifetime. 

For simplicity we ignore the effects of mass-loss through stellar winds, such that every system remains at its initial mass-ratio, $q$, until mass-transfer occurs (which may be either conservative or non-conservative). Also, given the fact that the stellar mass-luminosity relation is very steep, we define each binary system by its most luminous component, so that each binary can be assigned an equivalent single-star mass. To facilitate comparison with open cluster observations, our synthetic population is assumed to be coeval.

Other properties of our population are not designed to maximise the efficiency of Be star formation, and are more or less standard in binary evolution calculations. We denote the initial masses of the initially more massive star as $M_1$, the initially less massive as $M_{2,i}$ and define the initial mass-ratio, $q$, as 
\begin{align}
q=\frac{M_{2,i}}{M_1} ,  \label{Eq:q_def}
\end{align}
such that 
\begin{align}
0 < q \leq 1 .
\end{align}

We consider a population of binary stars where the distribution of initial primary mass follows a power law like
\begin{align}
\xi(M_1) = \xi_0 M_1^{\alpha}, \label{Eq:m1dist}
\end{align}
and the distribution of initial mass-ratios is described similarly as
\begin{align}
f(q) = f_0 q^{\kappa}, \label{Eq:qdist}
\end{align}
where $\xi_0$ and $f_0$ are normalising constants to ensure that the integral over the whole parameter space is unity (as befitting a probability-density function). For example, the value of $f_0$ is easily computed as 
\begin{align}
f_0 = \frac{\kappa +1}{1- q_{min}^{\kappa +1}}, \label{Eq:fo_def}
\end{align}
where $q_{min}$ is the minimum mass-ratio in our population and will be nominally set to $q_{min}=0.1$ to match the observing campaigns of \citet{2012Sci...337..444S,2013A&A...550A.107S}. It is assumed that systems born with mass-ratios smaller than this value are likely to be unstable and merge either during their formation or early in their evolution and hence are not considered. 

Mass gain of the accretor shall be parameterised by assuming that a total mass of $\Delta M$ is accreted, giving the relation between final and initial masses of the accretor as 
\begin{align}
M_{2,f} = M_{2,i} (1 + \Delta M/M_{2,i}), \label{Eq:deltaM}
\end{align}
with $\Delta M/M_{2,i}$ being a free parameter. 


Our assumptions on the population are summarised in the list below.
\begin{enumerate}
\item initial binary fraction is 1
\item every system will undergo stable Case B mass-transfer and form a Be star, irrespective of period or mass ratio
\item once a primary star leaves the main-sequence a Be star is immediately formed
\item once a Be star is formed, it remains so for the rest of its lifetime
\item the accretor star gains mass $\Delta M$, with the relative mass-gain $\Delta M/M_{2,i}$ being a free parameter. 
\item the effects of wind mass-loss are ignored so that a system remains at its initial mass-ratio until mass-transfer occurs.
\item the distribution of initial primary masses follows a power law, $\xi(M_1) \propto M_1^{\alpha}$
\item the distribution of initial mass-ratios follows a power law, $f(q) \propto q^{\kappa}$
\item only considered are binaries with a mass-ratio greater than $q_{min}$, which is set to 0.1.
\item when both stars are hydrogen-burning, the luminosity of a binary system is given by that of the primary. When the primary has evolved off the main-sequence, the luminosity of the system is naturally that of the secondary.
\end{enumerate}


According to our assumptions, every secondary star with a post-main-sequence companion is a Be star, meaning the number of Be stars with given mass $M$ is 
\begin{align}
n(\textrm{Be}) = n(M_{2,f} = M \, \& \, M_1 > M_{TO}),
\end{align}
where $M_{TO}$ is the turn-off mass of our coeval population. In our model the number of non-Be stars is given by the number of primaries at a given mass. The Be fraction, $\phi_{\textrm{Be}}(M)$ shall be defined as the number fraction of Be stars to all stars at a given mass. In a coeval population this becomes 
\begin{align}
\phi_{\textrm{Be}}(M) &= \frac{n(M_{2,f}=M \,\& \, M_1 >M_{TO})}{n(M_{2,f}=M \, \& \, M_1 >M_{TO}) + n(M_1=M)}  \\
&=  \left[ 1+ \frac{n(M_1=M)}{n(M_{2,f}=M \, \& \, M_1 >M_{TO}) }\right] ^{-1} .
\end{align}
In our model, a Be star's mass is related to its initial mass, $M_{2,i}$, and the relative mass-gain, $\Delta M/M_{2,i}$ via Eq. \ref{Eq:deltaM}, such that the expression above becomes
\begin{align}
\phi_{\textrm{Be}}(M) &=  \left[ 1+ \frac{n(M_1=M)}{n(M_{2,i}=\frac{M}{1+\Delta M/M_{2,i}} \, \& \, M_1 >M_{TO}) }\right] ^{-1} .
\end{align}
With the aid of Eqs. \ref{Eq:q_def} and \ref{Eq:deltaM}, the condition 
\begin{align}
M_1> M_{TO}
\end{align}
can be rewritten as 
\begin{align}
q&< \frac{M_{2,i}}{M_{TO}}, 
\end{align} 
leading to
\begin{align}
q&< \frac{M_{2,f}}{M_{TO}(1+\Delta M/M_{2,i})}.
\end{align} 
This results in
\begin{align}
\phi_{\textrm{Be}}(M) &=  \left[ 1+ \frac{n(M_1=M)}{n(M_{2,i}=\frac{M}{1+\Delta M/M_{2,i}} \, \& \,q< \frac{M}{M_{TO}(1+\Delta M/M_{2,i})}) }\right] ^{-1} .
\end{align}

To study coeval populations, a more convenient approach is to find the Be fraction as a function of the fractional main-sequence turn-off mass, $M/M_{TO}$. This produces the expression
\begin{align}
\phi_{\textrm{Be}}(M/M_{TO}) &=  \left[ 1+ \frac{n(M_1=M/M_{TO})}{n(M_{2,i}=\frac{M}{M_{TO}(1+\Delta M/M_{2,i})} \, \& \,q< \frac{M}{M_{TO}(1+\Delta M/M_{2,i})}) }\right] ^{-1}, \label{Eq:befrac_def2}
\end{align}
which shall be our basis for exploring the Be fraction in coeval populations.


To evaluate the Be fraction, it is necessary to find the relative numbers of primary stars to that of secondary stars at a given mass. We may write that the number of primary stars with a given mass, $n(M_1 )$, is the integral of the primary mass distribution across an infinitesimally small mass range, $dM_1$, multiplied by the total number of stars in the population, $n_{tot}$, as
\begin{align}
n(M_1 ) = n_{tot} \xi({M_1}) d M_1 = n_{tot} \xi_0 M_1 ^{\alpha} d M_1. \label{Eq:6}
\end{align}

To tackle the number of secondary stars at a given mass is sightly more involved as we do not have directly the distribution of secondary masses, instead it is inferred from the primary mass and mass-ratio distributions. First consider a population in which there exists only a single mass-ratio, $q_0$, ie. the mass-ratio distribution is a delta-Dirac function. If one is interested in the number of secondary stars with initial mass $M_{2,i}$, one must count the number of primaries with mass $M_{2,i}/ q_0$, so we have 
\begin{align}
n(M_{2,i} \, \& \, q_0) =n_{tot} \xi\left(\frac{M_{2,i}}{q_0}\right) d\left(\frac{M_{2,i}}{q_0} \right),
\end{align}
with $d\left(\frac{M_{2,i}}{q_0} \right)$ representing an infinitesimally small change in $\frac{M_{2,i}}{q_0}$.

Any distribution may be expressed as an infinite sum of appropriately weighted delta-Dirac distributions, with the weighting coming from the probability-density function. Therefore for the general case we have 
\begin{align}
n(M_{2,i}) = n_{tot} \int_{q_{min}} ^ {1} f(q) \xi\left(\frac{M_{2,i}}{q}\right) d\left(\frac{M_{2,i}}{q} \right) dq. \label{Eq:1.1}
\end{align}
It is then clear that the limits of the integral above place constraints on the initial mass-ratios counted. The number of systems with a given initial secondary mass $M_{2,i}$ and initial mass-ratios between $q_{min}$ and $q_{max}$ can thus be written as 
\begin{align}
n(M_{2,i} \, \& \, q_{min} < q < q_{max}) = n_{tot} \int_{q_{min}} ^ {q_{max}} f(q) \xi\left(\frac{M_{2,i}}{q}\right) d\left(\frac{M_{2,i}}{q} \right) dq, \label{Eq:1}
\end{align}
 
The differential $d\left(\frac{M_{2,i}}{q} \right)$ in Eq.\ref{Eq:1} is quite cumbersome so we chose to let 
\begin{align}
r= \frac{M_{2,i}}{q}.
\end{align}
We may now write
\begin{align}
n(M_{2,i} \, \& \, q_{min} < q < q_{max}) = n_{tot} \int_{q_{min}} ^ {q_{max}} f(q) \xi(r) dr dq. \label{Eq:4}
\end{align}
We have 
\begin{align}
M_{2,i} = qM_1,
\end{align}
thus
\begin{align}
dM_{2,i} = q dM_1 + M_1 dq.
\end{align}
As we are interested in the number of secondary stars at a fixed mass, {$d{M_{2,i}} =0$}, so 
\begin{align}
dq= -\frac{q}{M_1} dM_1. \label{Eq:2}
\end{align}
Differentiating $r$ gives
\begin{align}
dr= \frac{-M_{2,i}}{q^2} dq. \label{Eq:3}
\end{align}
Combining Eqs. \ref{Eq:2} and \ref{Eq:3} results in
\begin{align}
dr = \frac{1}{q}\frac{M_{2,i}}{M_1} dM_1.
\end{align}
Inserting this into our expression for $n(M_{2,i} \, \& \, q_{min} < q < q_{max})$ (Eq. \ref{Eq:4}) gives
\begin{align}
n(M_{2,i} \, \& \, q_{min} < q < q_{max}) &= n_{tot} \int_{q_{min}} ^ {q_{max}} f(q) \xi\left(\frac{M_{2,i}}{q}\right) \frac{1}{q}\frac{M_{2,i}}{M_1} dM_1 dq.\label{Eq:5}
\end{align}
We now divide Eq. \ref{Eq:6} by Eq. \ref{Eq:5} leaving 
\begin{align}
\frac{n(M_1)}{n(M_{2,i} \, \& \, q_{min} < q < q_{max})} = \frac{M_1 \xi(M_1)}{M_{2,i} \int_{q_{min}} ^ {q_{max}} f(q) \xi\left(\frac{M_{2,i}}{q}\right) \frac{1}{q} dq} \label{Eq:7}.
\end{align}
When the distributions for initial primary mass and mass-ratio, Eqs. \ref{Eq:m1dist}\, and \ref{Eq:qdist}, are inserted, Eq. \ref{Eq:7} simplifies further to 
\begin{align}
\frac{n(M_1)}{n(M_{2,i} \, \& \, q_{min} < q < q_{max})} =\left(\frac{M_1}{M_{2,i}}\right)^{\alpha+1} \frac{1}{\int_{q_{min}} ^ {q_{max}} f_0 q^{\kappa -\alpha -1}  dq}  \label{Eq:finalfraction}.
\end{align}
This result may be readily checked against Monte-Carlo sampling of the primary mass and mass-ratio distributions.

Equation \ref{Eq:finalfraction} can be used to directly determine the Be fraction (Eq. \ref{Eq:befrac_def2}) by setting $M_1=M/M_{TO}$ , $M_{2,i}=\frac{M}{M_{TO}(1+\Delta M/M_{2,i})}$ and $q_{max}=\frac{M}{M_{TO}(1+\Delta M/M_{2,i})} $. This leaves

\begin{align}
\phi_{Be}(M/M_{TO})  = \Bigg[ 1+    \frac{( 1+ \Delta M /M_{2,i})^{(\alpha +1)}  }{ \int_{q_{min}} ^ {\frac{M_{}}{( 1+ \Delta M/M_{2,i})M_{TO}}} f_0 q^{\kappa - \alpha-1 }  dq}\bigg]^{-1}, \label{Eq:OCfinal}
\end{align}
where the integral has a simple analytic solution. Eq. \ref{Eq:OCfinal} describes the Be star fraction as a function of the fractional turn-off mass, $M/M_{TO}$ for our model open cluster. As the mass-dependence in  Eq. \ref{Eq:OCfinal} is expressed by the fractional turn-off mass, is not necessary to specify the turn-off mass.

\subsection{Limits for remaining parameters \label{sec:paramchoice}}
All that remains is to explore Eq. \ref{Eq:OCfinal} in a suitable parameter space. The parameters we have are $\alpha$, $\kappa$ and $\Delta M/M_{2,i}$, the primary mass distribution exponent, initial mass-ratio distribution exponent and relative accretor mass-gain respectively.

The canonical value for the initial-mass-function (IMF) exponent, $\alpha$, is given by the Salpeter IMF, $\alpha= -2.35$ \citep{1955ApJ...121..161S}. However recent observations of young stars in the 30 Doradus starburst region suggest instead $\alpha=-1.90 ^{+0.37} _{-0.26}$ \citep{2018Sci...359...69S}. Similarly in the R136 star forming region an exponent of $\alpha=-2.0 {\pm0.3}$ was found \citep{2020MNRAS.tmp.2627B}. On the other hand it has also been proposed that the IMF follows an even steeper law with $\alpha=-2.7$ \citep{1986FCPh...11....1S}. Therefore we consider the range $-1.9 < \alpha < -2.7$ .

Observations of Galactic O-type stars show that the mass-ratio distribution follows a power law with exponent $\kappa=-0.1 \pm 0.6$ \citep{2012Sci...337..444S} for $0.1< q< 1$. In the Large Magellanic Cloud, the mass-ratios of massive binaries appear to be distributed differently with $\kappa = -1.0 \pm 0.4$ again in the range $0.1< q< 1$ \citep{2013A&A...550A.107S}. There are many claims that mass-ratios of binaries favour either low values  \citep{1990MNRAS.242...79T,1991MNRAS.250..701T,1991PhDT.......257H} or follow a uniform distribution\citep{2007ApJ...670..747K,2007A&A...474...77K}. In light of these findings we shall consider $\kappa$ values in the range $-1 < \kappa  < 0$.

Estimates of the accretor mass-gain, $\Delta M/M_{2,i}$, obtained by demanding that mass-transfer stops once the mass-gainer reaches critical rotation tell us that $\Delta M/M_{2,i}$ is at the very most 0.1 and in most cases around 0.02, depending on the angular momentum content and physical structure of the mass-gainer before accretion \citep{1981A&A...102...17P,2005A&A...435.1013P,2020ApJ...888L..12W}. It has been found that around 70\% of the mass leaving the donor must be ejected from the system to explain observed distributions of Be star masses in Be X-ray binaries  \citep{2020arXiv200300195V}. However it must be noted that because it is expected that up to 90\%  of massive binary systems are broken apart by a supernova kick \citep{1995MNRAS.274..461B},  Be X-ray binaries represent a small fraction of the population, and hence may well contain strong biases. Furthermore, it is believed that mass-transfer must be highly non-conservative to explain observed populations of Wolf-Rayet O-star binaries \citep{2005A&A...435.1013P,2016ApJ...833..108S}. On the other hand, several systems exist which show evidence of near-conservative mass-transfer having taken place \citep{2007A&A...467.1181D,2018A&A...615A..30S,2021A&A...645A..51B}. To fully explore the effects of mass-transfer efficiency on Be star populations, we take the full range $0< \Delta M/M_{2,i} < 1 $.

\section{Results \label{sec:results}}

\begin{figure*}
	\includegraphics[width=1.0\linewidth]{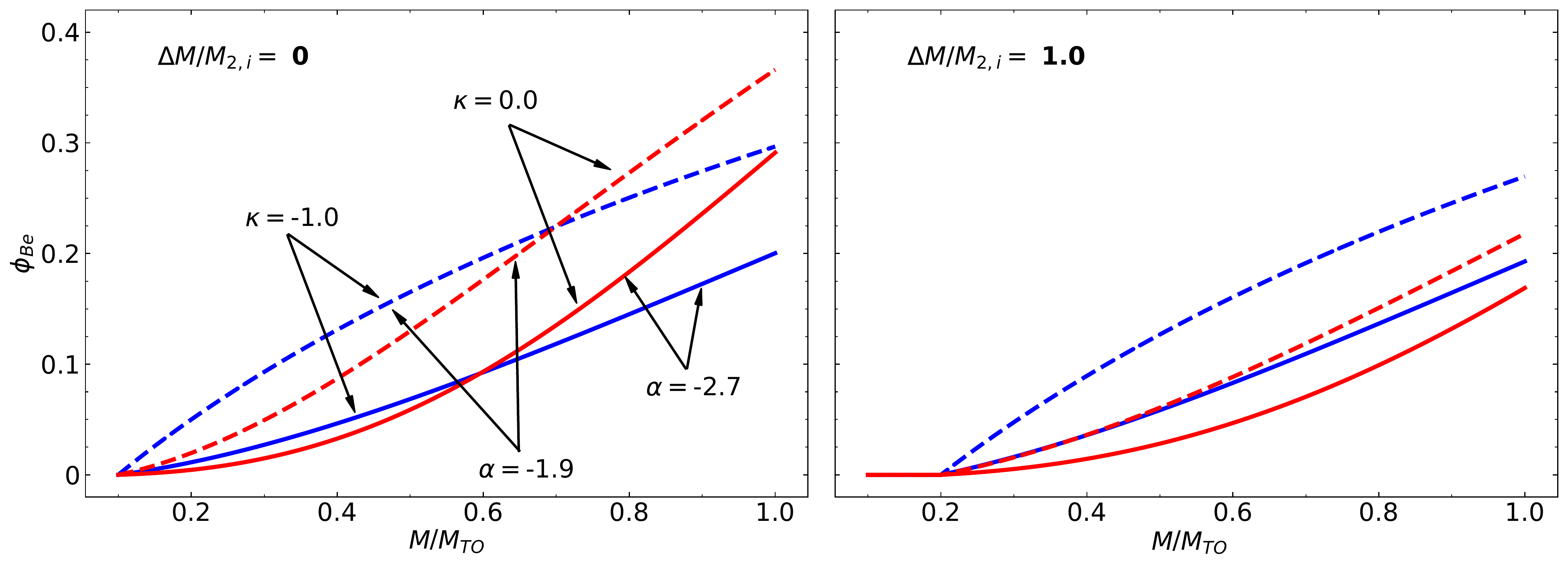}
	\centering
	\caption{The maximum Be fraction, $\Phi_{Be}$, in a coeval population as defined by Eq.\, \ref{Eq:OCfinal} plotted as a function of fractional main-sequence turn-off mass, $M/M_{TO}$ for varying parameters. The left and right Panels show $\Delta M/M_{2,i}=$ 0, and 1 respectively. The colour of the lines represents differing $\kappa$ values with red being $\kappa=0$ and blue $\kappa=-1$. Dashed lines show $\alpha=-1.9$ and solid lines $\alpha=-2.7$ as indicated by the annotations. }
	\label{fig:OC1} 
\end{figure*}

The results of Eq.\, \ref{Eq:OCfinal} are plotted in Fig.\, \ref{fig:OC1} for the extremal parameters outlined in Sect.\,\ref{sec:paramchoice}.  The primary mass distribution affects the absolute numbers of Be stars because for a shallower distribution ($\alpha$ closer to 0), there is an abundance of massive binaries, such that many systems contain post main-sequence primaries and therefore the number of Be stars increases. Conversely when $\alpha << 0$, the population contains fewer primaries of mass greater than the turn-off mass and the Be count decreases.

The effect of the mass-ratio distribution can be understood by considering a population with a high value of $\kappa$ such that secondary stars have a similar mass to their companion. In this case, when the primary leaves the main-sequence, the secondary will be rather evolved, and hence most Be stars will be found near the turn-off. On the other hand in a population with a low $\kappa$, the opposite is true; the secondary stars will have low masses compared to the turn-off mass and Be stars will be more evenly distributed along the main-sequence, as seen in Fig.\, \ref{fig:OC1}.

Figure\, \ref{fig:OC1} shows how a varying mass-gain changes the Be count, with accretors that gain more mass producing fewer Be stars. This can be understood by considering a Be star of mass 0.9$M_{TO}$ that is produced by inefficient mass-transfer. For the primary to exceed the turn-off mass, the initial mass-ratio of the system must be less than 0.9. Now if this star had gained 0.1$M_{TO}$, the initial mass would be 0.8$M_{TO}$ and the initial mass-ratio must be less than 0.8. Therefore, mass gain restricts the number of systems that are able to produce Be stars of a given mass, and low mass-transfer efficiency leads to higher numbers of Be stars being produced at a given mass.

It is seen in Fig.\, \ref{fig:OC1} that the largest Be fractions are produced when mass-transfer is inefficient ($\Delta M/M_{2,i}=0$) and the initial mass function is shallow. The mass-ratio distribution then tunes the distribution of Be stars along the main-sequence. Therefore it is judged that the most Be stars are produced with the parameters $\Delta M/M_{2,i}=0$, $\alpha=-1.9$ and $-1 < \kappa <0$. Depending on the chosen parameters, the maximum Be fraction is in the range 0.2-0.35 near the main-sequence turn-off.

\section{Comparison to observations \label{sec:obscomp}}

To contextualise to our results, we attempt here a comparison with observations using high quality Hubble Space Telescope photometry of young Small and Large Magellanic Cloud open clusters in which Be stars are revealed as bright objects in a narrow-band filter centred on H$\alpha$ \citep{MiloneObs}. {Photometry was performed with Hubble wide-band filters $F814W$ and $F336W$ and the narrow band $F656N$ filter, allowing one to produce colour-magnitude diagrams in which Be stars are identified from H$\alpha$ photometry.} 

{
 As many spectroscopically confirmed Be stars in NGC 330 are bright in H$\alpha$ \citep{2020A&A...634A..51B}, we judge H$\alpha$ emission to be a good proxy for Be stars. It is possible for the accretion discs of Algol-type binaries to exhibit H$\alpha$ emission \citep{1989SSRv...50....9P} however, such systems are expected to contribute around 3\% to the total population (\citealt{2014ApJ...782....7D}, Sen et al. in prep.).  Furthermore, some field stars may be  H$\alpha$ emitters, with \citet{MiloneObs} noting that no more than one-tenth of stars in the cluster field are suspected field stars. Field stars will also contaminate the population of stars not emitting in H$\alpha$, therefore their presence is not expected to significantly alter the relative fractions of H$\alpha$ emitters and non-emitters. 
}

Be star fractions have previously been measured as a function of magnitude \citep{1999A&AS..134..489K,MiloneObs,2020A&A...634A..51B}. However we find it worthwhile to repeat this exercise, including several factors which were previously overlooked.

\subsection{Counting Be stars}

Our goal is to measure the observed Be fraction as a function of mass along the main-sequence of an open cluster. To do this we must note the two major differences between a Be star and a "normal" B star; fast rotation and the presence of a decretion disc. Due to the effect of the centrifugal force, a fast-rotating star suffers from reduced effective gravity at the equator, and according to the von Zeipel theorem \citep{1924MNRAS..84..665V}, this results in a lower effective temperature. Therefore fast-rotating stars are cooler and redder than their non-rotating counterparts. Furthermore, light from a Be star consists of radiation from the star itself and also light from the decretion disc. Typically the average temperature of the disc is around 70\% that of the star's effective temperature \citep{2009ApJ...699.1973S}, and so the disc is expected to emit mostly in visible and infra-red wavelengths. 

In a colour-magnitude diagram, the magnitude in a red filter is plotted on the y-axis and a colour defined by the blue and red filter (B - R) on the x-axis. When a star becomes brighter in the red filter, it will therefore move to the right and upwards in the colour-magnitude diagram. This effect means that to count the Be stars as a function of mass, we must do so in bins that are sloped with respect to the x-axis. The gradient of this slope depends on how much redder a near-critically rotating star is than a slow rotator at the same mass, and on how much light the decretion disc radiates. 

As no reliable numerical models exist of stars rotating at the critical velocity, we shall adopt a simple model to relate the luminosity and temperature of a critical rotator to an equivalent non-rotating star. After having been spun up, a star will change its shape, becoming oblate. At the same time we do not expect a great difference in luminosity between a star before and after the spin-up. This is because stars are generally very centrally condensed, such that the centrifugal force is small compared to gravity in the regions where nuclear burning occurs, meaning that (excluding the effects of rotational mixing) central temperatures and thus luminosities are not very sensitive to rotation, in agreement with models \citep{2011A&A...530A.115B,Paxton2019}. Following the Steffan-Boltzmann law, because of the increased surface area of a critical rotator, the effective temperature decreases. Using the Roche model (see Appendix \ref{App2}) , one can show that a critically rotating star has a surface area of approximately $1.58 \times 4\pi R_p ^2$, with $R_p$ being the polar radius, which corresponds to a decrease in effective temperature by a factor of ${1.58^{-\frac{1}{4}}}\approx 0.89$. Knowing this, we can construct isochrones describing the intrinsic properties of critically rotating stars from non-rotating isochrones. 

A further complication that is brought about by gravity darkening is that a fast rotating star appears cooler and dimmer when viewed equator-on as compared to pole-on. Assuming a random orientation of the inclination axis, the mean value of the sine of the inclination angle is $\pi /4$, corresponding to a mean inclination angle of 51.8 \si{\degree}. To take into account the mean effect of gravity darkening, we employ the model of \cite{2011A&A...533A..43E} as implemented in MESA \citep{Paxton2019}. Here, the projected luminosity and effective temperature, $L_{\textrm{proj}}$, $T_{\textrm{eff,proj}}$ are related to the intrinsic luminosity and effective temperature, $L$, $T_{\textrm{eff}}$ by 
\begin{align}
L_{\textrm{proj}} &= C_T(\omega , i) L \\
T_{\textrm{eff,proj}} &= C_L(\omega , i) T_{\textrm{eff}} ,
\end{align}
with $C_T$ and $C_L$ depending on the fraction of critical velocity, $\omega$ and inclination angle $i$.

\begin{figure}
	\includegraphics[width=1.0\linewidth]{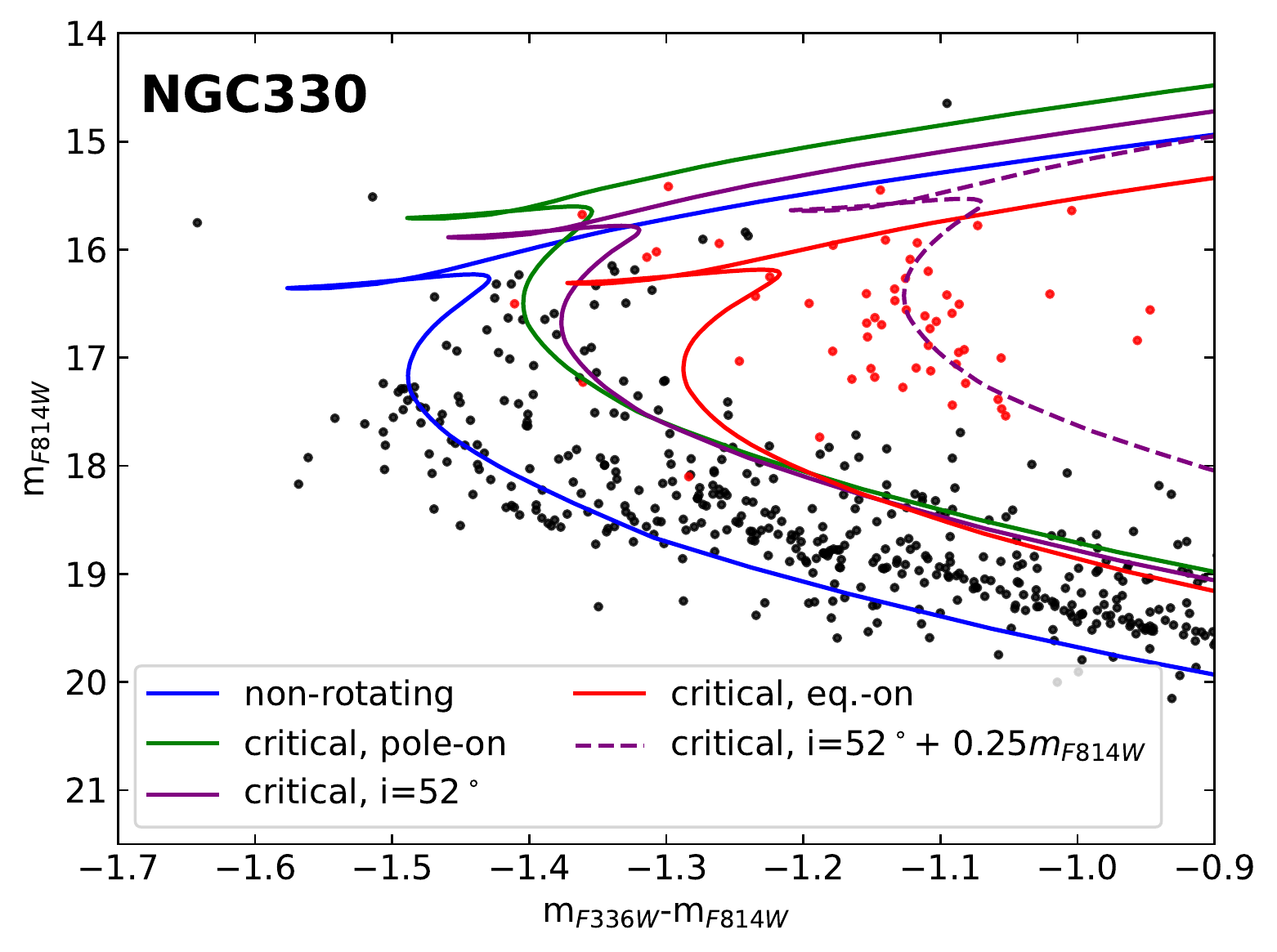}
	\centering
	\caption{Colour-magnitude diagram of NGC 330 focused on the turn-off region with H$\alpha$ emitters marked in red. An isochrone of non-rotating stars is plotted in blue (see App. \ref{App:isochrones} for model details). Green and red isochrones depict critically rotating stars viewed pole-on and equator-on respectively, as derived from a simple model of critical rotators (see App. \ref{App2}). The solid purple isochrone represents critically rotating stars viewed at the mean inclination angle when the rotation axis is randomly oriented (51.8\si{\degree}) and for the dotted purple 0.25\meightonefour has been added to simulate the decretion disc. The isochrone age is 30 Myr, distance modulus $\mu$ = 18.8\,mag and reddening of E(B-V)=0.1\,mag. Data from \citet{MiloneObs}  }
	\label{fig:disc} 
\end{figure}

The temperatures and luminosities of critically rotating stars are found by using the coefficients $C_T(\omega=1, i=51.8 \si{\degree})= 1.02$ and $C_L(\omega=1, i=51.8 \si{\degree})=1.22$. It is a rather curious feature of the gravity darkening model that at the mean inclination, the coefficients exceed unity, meaning the average effect of gravity darkening is not darkening at all, but brightening. Finally, by interpolating tables of synthetic stellar spectra \citep{2016ApJ...823..102C} to produce magnitudes in Hubble filters, we are able to produce an isochrone of critical rotators, as shown in Fig. \ref{fig:disc}.



The contribution of a Be star's disc to its total flux is more difficult to {assess}. It has been noticed that a loss of spectral emission features in certain Be stars coincides with a dimming of around 0.3-0.5 magnitudes in the R and V filters \citep{2012ApJ...744L..15C,2017AJ....153..252L,2018MNRAS.476.3555R}. If the loss of emission features is interpreted as the disappearance of the disc, one can take this change in brightness to equal the flux contribution of the disc. By comparing the colour of our isochrones with the colours of Be stars in NGC 330, we can assess how much the Be disc shines, as in Fig. \ref{fig:disc}. After assuming that the disc shines in the \eightonefour filter but not in the \threethreesix filter, we find a reasonable fit to the H$\alpha$ emitters when a disc brightness of 0.25\meightonefour is adopted, as shown by the solid and dashed purple lines in Fig. \ref{fig:disc}.  

For NGC 330, in the regions of the colour-magnitude diagram containing Be stars, we find stars of equal mass on the non-rotating and Be star isochrones to be connected by lines of gradient $\frac{-d{m}_{\textrm{F814W}} }{d(m_{\textrm{F336W}}-{m}_{\textrm{F814W}})}=$ 2. For NGC 2164, again assuming a constant disc magnitude of 0.25\meightonefour, the gradient is found to be 1.8. These differing values are caused by the ways in which stellar spectra, and hence magnitude in a given filter, vary with luminosity and effective temperature.



\begin{figure*}
	\includegraphics[width=1.0\linewidth]{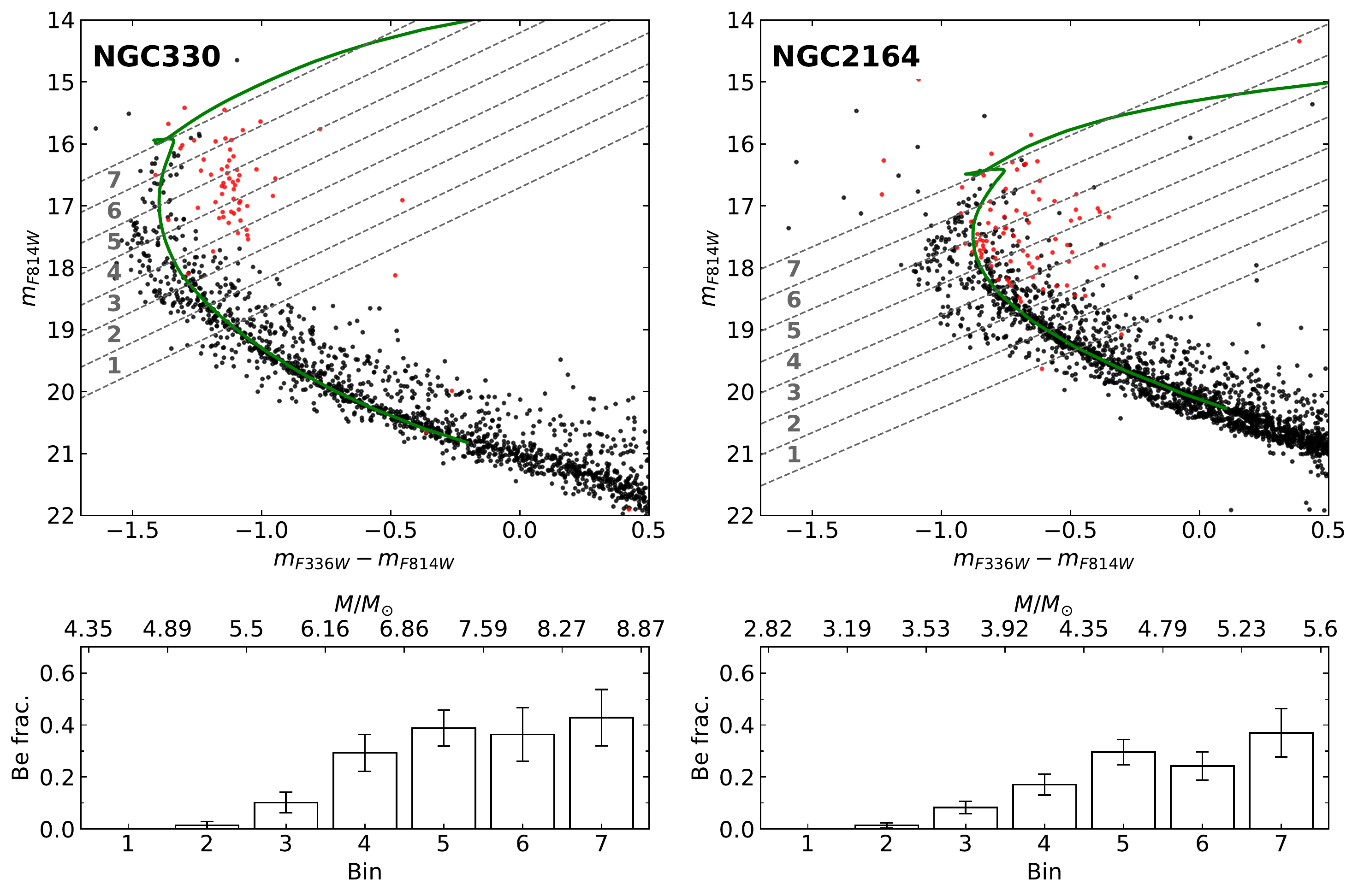}
	\centering
	\caption{Colour-magnitude diagrams with isochrone fits and Be star counts for Small Magellanic Cloud cluster NGC 330 (left) and Large Magellanic Cloud cluster NGC 2164 (right). H$\alpha$ emitters are marked in red. Bottom panels show the Be fraction counted in bins as defined in the top panels with the errors given by the binomial counting error. The bins have a gradient of 2.0 and 1.8 for NGC 330 and NGC 2164 respectively. Mass values associated with the bins are provided by the isochrone fit. For both clusters the isochrone depicts stars with initial rotation equal to 0.6 ${v_{\textrm{rot}}}/{v_{\textrm{crit}}}\thinspace$. For NGC 330 the isochrone age is 30 Myr, distance modulus $\mu$ = 18.8\,mag and reddening of E(B-V)=0.1\,mag. For NGC 2164 the age is 80 Myr, $\mu$= 18.3\,mag and E(B-V)=0.12\,mag. Data from \citet{MiloneObs}  }
	\label{fig:obs} 
\end{figure*}

Figure \ref{fig:obs} shows the colour-magnitude diagrams of NGC 330 and NGC 2164 with the Be fraction as counted in slanted bins with gradients of 2.0 and 1.8 respectively. It is noted that as compared to counting the Be fraction in bins of constant \meightonefour magnitude (ie. horizontal bins) the values measured here are lower because in the horizontal bin case, one is counting B stars with a higher mass than the Be stars in the same bin. According the initial-mass-function, the higher mass stars are less populous and hence the Be fraction increases solely because there are fewer B stars being counted. 


We use isochrones of rotating single stars based on an extended model grid of \citet{2019A&A...625A.132S} (see Appendix \ref{App:isochrones} for a thorough description) with an initial rotation rate of \Vcritfrac = 0.6 to assign mass-ranges to each bin, so that the Be fraction can be evaluated as a function of mass. The bins are placed so that the outer edge of the last bin is at the point where hydrogen has been exhausted in the stellar core. The value of \Vcritfrac = 0.6 is chosen as suggested by \citet{2019ApJ...887..199G} and Wang et al. (in prep.), and produces equatorial rotation velocities that are in broad agreement with spectroscopic observations \citep{2013A&A...550A.109D,2018AJ....156..116M,2019ApJ...876..113S,2020MNRAS.492.2177K}. The reddening and distance modulus values are tailored to give the best fit to the cluster and are in good agreement with previous isochrone fittings for these clusters \citep{MiloneObs}. The isochrone fits are shown in Fig \ref{fig:obs}.

\begin{figure}
	\includegraphics[width=1.0\linewidth]{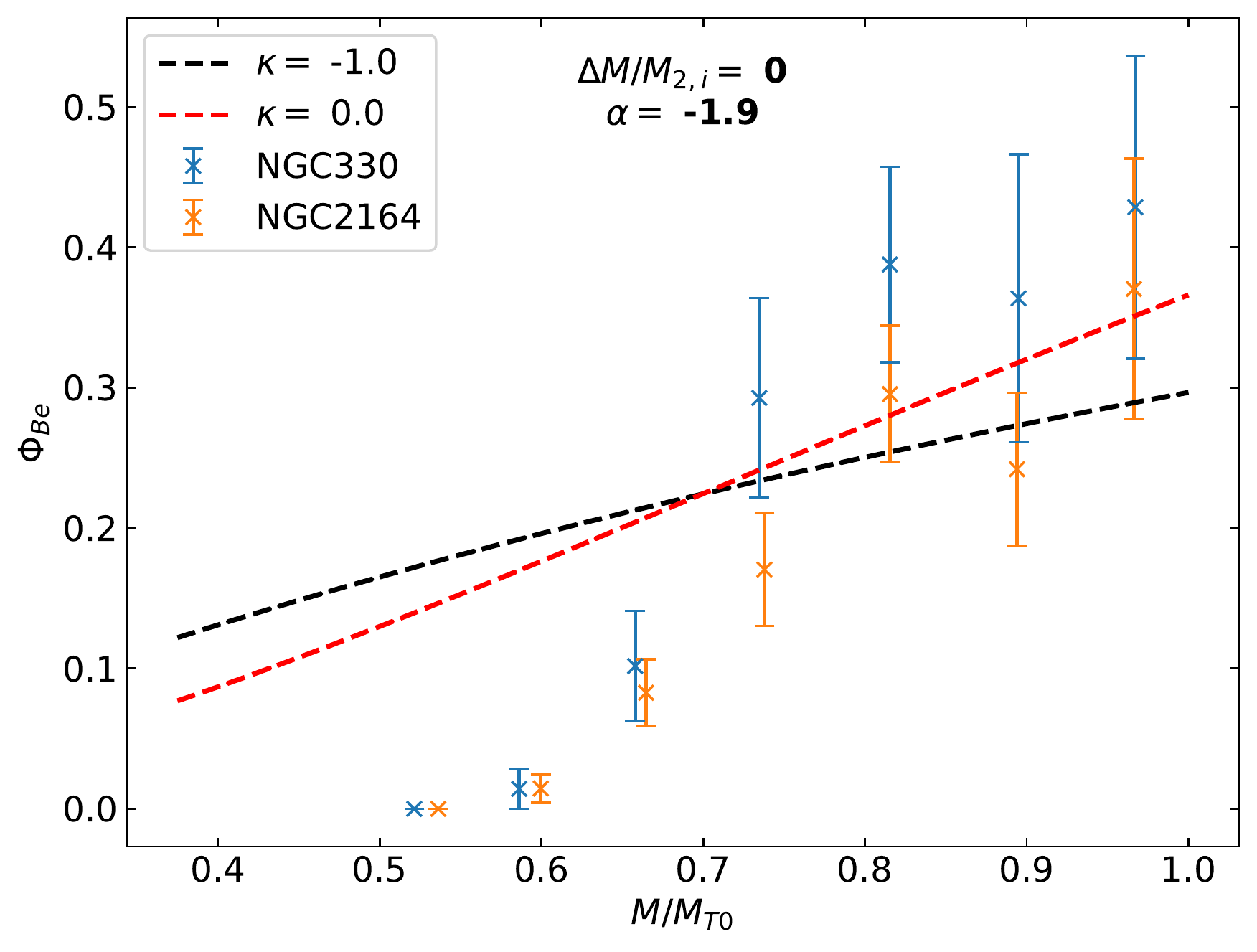}
	\centering
	\caption{Comparison between theory and observations. Be fraction as a function of fractional main-sequence turn-off mass in NGC 330 and NGC 2164 as shown in Fig. \ref*{fig:obs}. Dashed lines show theoretical upper limit given by Eq. \ref{Eq:OCfinal} with $\alpha=-1.9$, $\Delta M/M_{2,i}=0$ and $\kappa=-1.0,0$ (see Fig. \ref{fig:OC1}) as given by legend.}
	\label{fig:comp1} 
\end{figure}

The isochrones allow us to measure the turn-off mass and the masses associated with each bin, thereby a direct comparison between the theory presented in Sec. \ref{sec:results} and observations is possible. Figure \ref{fig:comp1} shows this comparison, with counting uncertainties on the Be fraction given by the standard error, $\sigma$, assuming a binomial distribution as 
\begin{align}
\sigma =\sqrt{\Phi (1-\Phi)/N},
\end{align}
with $\Phi$ being the measured Be fraction and $N$ the total number of stars in a given bin.

We find that despite the two clusters being of different metallicities and ages, they seem to have similar Be fractions as a function of relative turn-off mass. This may be an indication that whatever the dominant Be production channel is, it is universal.

In both clusters the Be fraction steadily increases from zero to around 0.4  in the range 60-80\% of the turn-off mass. Near the turn-off, the Be fraction is found to be approximately 0.4 with significant counting uncertainty due to the relatively small numbers of stars occupying this region. Taking into account these uncertainties, it is seen that  our upper limit can describe the numbers of Be stars in the upper part of the main-sequence. It is important to note that because of the difficulty in performing an isochrone fit, the Be fractions near the turn-off are particularly uncertain, with a small change in the isochrone fit resulting in a large change in the measured Be fraction (see Sec. \ref{sec:uncertainties} for a quantitative discussion). Therefore, despite the measured Be fraction in NGC 330 at times exceeding our upper limit, it is reasonable to conclude that the upper limit does provide a reasonable fit to the Be star numbers near the turn-off. However it does fail to explain the lack of Be stars below $M/M_{TO}\approx 0.7$. This may be the result of certain systems not forming Be stars but instead merging, as shall be discussed in the next section.





\section{Inferring the initial conditions for stable mass-transfer \label{sec:stable_conditions}}

The observations presented in Sec. \ref{sec:obscomp} show that our upper limit can approximately describe the numbers of Be stars near the turn off, but fails to reproduce the Be sequence's sharp cut-off. Here, we shall investigate how our prescription will need to be changed in order to reproduce this feature.

In reality, not every binary system will undergo stable mass-transfer to form a Be star. For the specific case of the donor being in the Hertzsprung gap, as it is in Case AB or Case B mass-transfer, the mass-transfer proceeds at the Kelvin-Helmholtz (or thermal) timescale \citep{1997MNRAS.291..732T,2001A&A...369..939W}, meaning that if there is a large discrepancy in the Kelvin-Helmholtz timescales of the donor and accretor, the mass-transfer will become unstable and a common envelope situation will ensue, most likely leading to a stellar merger. 


To model the occurrence of mergers, it is often assumed in simplified binary evolution calculations \citep{PolsBinaryModels,2002MNRAS.329..897H,2015ApJ...805...20S}  that systems below a certain mass-ratio will merge, however such a simple criterion is unsuitable to reproduce the observations shown in Fig. \ref{fig:obs}. Equation \ref{Eq:OCfinal} gives the Be star fraction as an integral quantity, such that the Be star fraction at the main-sequence turn-off is the accumulation of systems with mass-ratios from $q_{min}$ to 1. This may be understood intuitively by noting that a Be star of mass near the main-sequence turn-off mass can originate from either an extreme mass-ratio system with a very massive primary, or from a system with mass-ratio close to unity. Therefore, when we demand that all systems below a given mass-ratio merge, we will naturally decrease the Be fraction at the turn-off, which we must avoid to retrieve high numbers of Be stars at the turn-off.

\begin{figure*}
	\includegraphics[width=\linewidth]{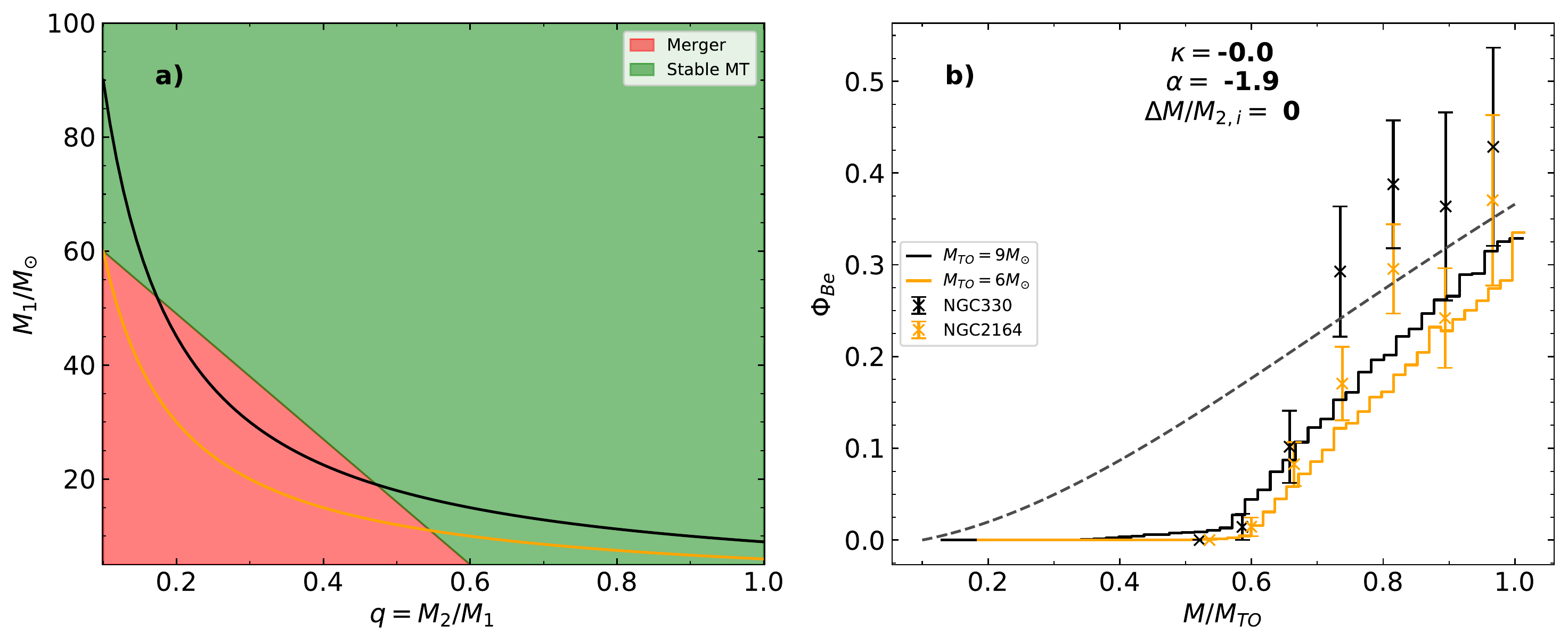}
	\centering
	\caption{\textbf{a)} Adopted region of stable mass-transfer in the primary mass-mass-ratio plane. Regions coloured red experience unstable mass-transfer and merge, while for green regions, mass-transfer is stable and a Be star is formed. The black and orange lines show systems with secondary masses of 9 and 6 \Msun respectively.
	 \textbf{b)} Results of a Monte-Carlo simulation showing the Be fraction, $\Phi_{Be}$, when the stable mass-transfer region in a) is applied. Binary systems have a flat mass-ratio distribution ($\kappa =0$), a primary mass distribution $\xi(M_1) \propto M^{-1.9}$ and we have assumed inefficient accretion ($\Delta M/M_{2,i}=0$). The black line shows a simulation with a turn-off mass of 9\Msol, and the orange line of 6\Msol. The dashed grey line shows the theoretical upper limit, as given by Eq. \ref{Eq:OCfinal}. Measured Be fractions of NGC 330 and NGC 2164 according to Fig. \ref*{fig:obs} are plotted as black and orange crosses respectively.}
	\label{fig:MC1} 
\end{figure*}

To keep the Be fraction near the turn-off high and produce a sharp break in the Be fraction at $M/M_{TO} \approx 0.7$, more sophisticated criteria are needed, namely with dependence on primary mass and mass-ratio. We propose that the systems most likely to suffer unstable mass-transfer are those with an extreme mass-ratio and low primary mass, as the components of such systems have the largest difference in Kelvin-Helmholtz timescales. This can be visualised in a grid of primary mass against mass-ratio, with the bottom corner consisting of systems that merge. In such a grid, systems with a fixed secondary mass are represented by parabolae, as depicted in Fig. \ref{fig:MC1}a. If the parabola representing a secondary with the turn-off mass can avoid the region containing merger progenitors, the Be fraction at the turn-off will remain close to the maximum theoretical prediction. Then as the secondary mass decreases, the parabolae will move into the corner with low mass-ratio and low primary mass and consequently the Be fraction will decrease.

To make a test of our hypothesis, we perform a Monte-Carlo simulation, whereby systems are picked randomly from given distributions of initial primary mass and initial mass-ratio. As before, we shall assume that mass-transfer is completely non-conservative ($\Delta M/M_{2,i}=0$). By choosing a turn-off mass, we can calculate the masses of Be stars in the simulation and therefore assess the Be fraction. The occurrence of mergers is decided using the stable mass-transfer region depicted in Fig. \ref{fig:MC1} a. The motivation for selecting this region will now be explained. Analysis of mass-transfer from giant donors \citep{2015MNRAS.449.4415P} has indicated that mass-transfer from Hertzsprung-gap stars is stable at mass-ratios greater than around 0.6. We will therefore assume that all systems with initial mass-ratios greater than 0.6 will undergo stable mass-transfer. The stability of mass-transfer is determined by the donor's reaction to mass-loss, where stars with radiative envelopes generally tend to contract as the envelope is being stripped \citep{1987ApJ...318..794H}. This is reversed for convective-envelope stars, which typically expand in response to mass loss \citep{1987ApJ...318..794H}. Stellar structure calculations suggest that stars with a mass greater than around 60\Msol spend very little time as red-giants, meaning that they mostly have radiative envelopes \citep{2019A&A...625A.132S,2020arXiv200611286K} such that mass-transfer is much more likely to occur when the donor has a radiative envelope. We therefore propose that mass-transfer will be stable for all systems with a primary mass exceeding 60\Msol. The region of instability is then defined by a linear interpolation between systems with $M_1 =60\Msol, q=0.1$ and $M_1=5\Msol, q=0.6$, as depicted in Fig. \ref{fig:MC1}a. We shall again assume that the orbital period plays no role in determining the stability of mass-transfer, hence allowing us to not specify an orbital period distribution.

Merger products change the distribution of masses in a population, hence affect the Be fraction as a function of mass. We will assume that Be stars are not merger products for two reasons. Firstly, it is believed that, although merger products are fast-rotators initially, while thermal equilibrium is returned, internal angular momentum redistribution causes a rapid spin-down \citep{2019Natur.574..211S}. What is more, stellar mergers may produce strongly magnetised stars \citep{2009MNRAS.400L..71F,2014MNRAS.437..675W,2019Natur.574..211S} which would further spin-down due to magnetic braking. Secondly, a merger between a star with a helium core and a main-sequence object will not produce a hydrogen burning star owing to the higher mean molecular weight and lower entropy of the helium-core star \citep{2012ARA&A..50..107L,2014ApJ...796..121J}. As observations \citep{MiloneObs} show Be stars to be concentrated on the main-sequence, we assume that Be stars are unlikely to be produced from the merging of two stars. For mergers, we assume the fraction of mass lost during the merging process to the total binary mass to be equal to 

\begin{align}
\mu_{loss} = \frac{0.3 q}{1+q^2},
\end{align}
\citep{2008A&A...488.1017G} which equates to between 2 and 15\% over the range $0.1<q<1$. As the mass lost during the merging process is assumed to be low, mergers will always have a mass exceeding the turn-off mass and will not affect the Be fractions on the main-sequence.

Figure \ref{fig:MC1} b shows the results of the simulation for clusters with turn-off masses of 9 and 6\Msol, which  roughly correspond to NGC 330 and NGC 2164, respectively. The chosen criteria have maintained a high Be fraction near the turn-off and also produced a sudden end to the Be-sequence, and provide a reasonable fit to the measured Be fractions in NGC 330 and NGC 2164. It is remarkable that such simple, although physically motivated, stable mass-transfer criteria can successfully reproduce the numbers of Be stars in the open clusters studied. Our empirical mass-transfer stability criteria could be tested in the next generation of detailed binary evolution models.

\section{Discussion \label{sec:disc}}
\subsection{Uncertainties \label{sec:uncertainties}}
\begin{figure*}
	\includegraphics[width=1.0\linewidth]{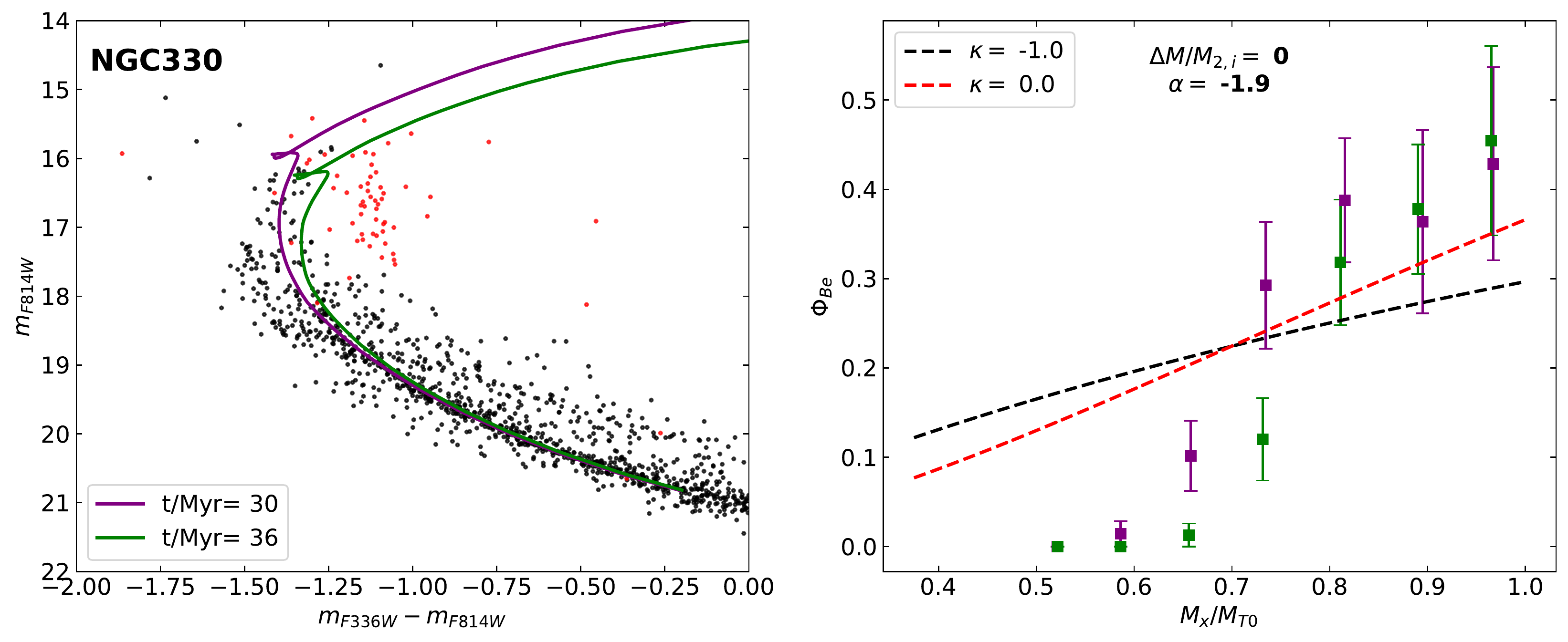}
	\centering
	\caption{Left panel: colour-magnitude diagram of NGC 330 with H$\alpha$ emitters marked in red. Isochrone fits with ages 30 and 36Myr are plotted in purple and green respectively. Both isochrones have initial rotation of \Vcritfrac =0.6, a distance modulus of 18.8\,mag and reddening of E(B-V)=0.1\,mag. 
	Right panel: Be fraction as a function of fractional turn-off mass as measured by the 30Myr isochrone (in purple) and the 36Myr isochrone (in green). Dashed lines show theoretical upper limit given by Eq. \ref{Eq:OCfinal} with $\alpha=-1.9$, $\Delta M/M_{2,i}=0$ and $\kappa=-1.0,0$ as given by legend. }
	\label{fig:uncert1} 
\end{figure*}

The largest uncertainty in our procedure comes from the isochrone fits. Most, if not all open clusters display an extended main-sequence turn-off, making the choice of a suitable isochrone age difficult. This is illustrated in Fig.\, \ref{fig:uncert1}, where isochrones of two different ages are fitted to NGC 330 and the Be fractions are evaluated. It is seen that a small variation in the adopted age can cause the Be count in some bins to vary by up to 0.2, with the end of the Be sequence being particularly affected. A similar sensitivity is also found for small differences in the distance modulus, reddening and isochrone rotation rates. From Fig.\, \ref{fig:uncert1}, it is judged that the uncertainty on the measured Be fractions is approximately 0.1 without including the counting error.

To measure the observed Be fraction as a function of mass, we must use slanted bins. In calculating the gradient of these bins, we have assumed that the Be star disc always adds 0.25\meightonefour to the magnitude of the star. This may be an over-simplification, with Be stars of differing mass or evolutionary status hosting relatively brighter or dimmer discs. Unfortunately this effect is difficult to observe and characterise and is also compounded by the fact that Be stars can display spectral and photometric variability \citep{2003PASP..115.1153P}. \citet{2009ApJ...699.1973S} report that the ratio of stellar effective temperature to mean disc temperature and infrared excess are indeed functions of spectral type.


Far older clusters, such as the 300Myr old NGC 1856 have much lower Be fractions than their younger counterparts \citep[see][Fig. 17]{MiloneObs}. Our simple model and mass-transfer stability criteria predict that the Be fraction does not vary strongly with turn-off mass and therefore is unable to explain the turn-off Be fraction in NGC 1856 of around 0.2. However this discrepancy may be partly explained by a changing binary fraction with mass, as it is known that more massive stars display a stronger preference for binary companionship \citep{2006A&A...458..461K,2009A&A...493..979K}, thus older clusters which contain fewer binaries will naturally have fewer Be stars. Another aspect behind the emission line phenomenon is the ionising power of the star, because to produce an emission line, the central Be star must ionise its decretion disc. Without sufficient ionising power, even if a decretion disc is present, no emission line will be observable and hence the star will seem ordinary. The ionising photon emission rate is known to be strongly dependant on effective temperature \citep{2003ApJ...599.1333S}, so at some limiting mass one would expect the central star to be unable to ionise a disc. This effect may play a role in lowering the Be fraction in older clusters and causing a dearth of Be stars at low magnitudes in the colour-magnitude diagram. Both of the clusters studied here have several stars that are very red, despite not being marked as H$\alpha$ emitters, which could in fact be such "dormant Be stars".

Lastly we note that in our work we have assumed that the properties of binary systems are distributed according to very simple laws. In reality however, the distributions may well be complex functions of one another, for example the mass-ratio distribution might be a function of the primary mass. The nature of these distributions is set by poorly understood binary star formation mechanisms as outlined by \citet{2020MNRAS.491.5158T}.



\subsection{Mass-transfer efficiency}

To construct our prediction of Be star fractions, we have assumed that no mass is accreted during mass-transfer and seen that this scenario fits observations reasonably well. It has been demonstrated that for efficient mass-transfer, the Be star fraction decreases, meaning that the theoretical framework presented here would not fit observations if mass-transfer were efficient. This leads us to propose that, if the binary Be formation channel is the dominant one, mass-transfer is, on average, far from conservative. 

Binary models with conservative mass-transfer predict Be stars to be blue stragglers, after having gained a lot of mass \citep{1997A&A...322..116V}. The observations presented in Fig. \ref{fig:obs} contradict this prediction, with the vast majority of Be stars lying either on the main-sequence or being slightly redder than it, strengthening our conclusion about mass-transfer being highly non-conservative.


\subsection{On the initial binary fraction}

In obtaining our results we have assumed an initial binary fraction of 1, which could be criticised as too extreme.
We have demonstrated that in a coeval population of binary systems, at most 30\% of systems are post-interaction binaries (see Sec. \ref{sec:results}). Pre-interaction systems would therefore make up no less than 70\% of this population. Dedicated models show that under the assumption of a constant star formation rate, $30^{+10}_{-15}$ \% of massive stars are the products of binary interaction \citep{2014ApJ...782....7D}, in broad agreement with this work. 

 Post-interaction binaries are either merger products, contain a relatively low mass post-main-sequence object (helium star, black-hole, neutron star or white-dwarf) with a main-sequence (possibly emission line) star or form a runaway star ejected from the binary orbit after a supernova. Such objects manifest themselves either as single stars, or would be difficult to detect as binaries \citep{2011IAUS..272..531D,2011BSRSL..80..543D}. Thus, even in a population whose initial binary fraction is 1, apparently single stars are present in the proportions described above.

By examining radial velocity variations of very massive stars, one may only measure the pre-interaction binary fraction (as supernova kicks are believed to disrupt almost all binary systems \citep{1995MNRAS.274..461B}), which has been observed to be around 0.7 for O-type stars \citep{2012Sci...337..444S}. We therefore argue that the initial binary fraction is certainly greater than the observed pre-interaction binary fraction, such that at this stage, we must remain open to the possibility that an initial binary fraction very close to one is indeed realised in nature.

\section{Conclusions \label{sec:conc}}

In light of various uncertainties plaguing binary evolution calculations, we have investigated whether binary evolution can possibly reflect the large numbers of Be stars observed in open clusters. Starting from the premise that any binary system, regardless of primary mass, orbital period or mass-ratio, will undergo stable mass-transfer to form a Be star, we have calculated a rigorous upper limit to the formation of Be stars through this channel. It has been demonstrated that such binary evolution does not allow more than around 30\% of stars to have been spun up though binary interaction and become emission line objects.

After using isochrone fits to assign stars in the colour-magnitude diagram masses, a count of the H$\alpha$ emitters in two open clusters reveals that for objects near the turn-off, our upper-limit provides a reasonable description of the numbers of Be stars, especially when uncertainties arising from the counting method are taken into account. The upper limit does however fail to describe the sudden decrease in Be fraction that both clusters exhibit at a mass approximately 70-80\% of the turn-off mass.

This problem can be rectified by assuming that systems of low mass-ratio and low primary mass merge. By adopting simple, although physically justified stable mass-transfer criteria, we have shown that a good fit to the observational data is produced by this postulate.

It has been demonstrated in a qualitative way that in coeval populations, a larger mass-gain of the donor results in a smaller Be fraction at a given mass. Given that the observed Be fractions are very close to our upper limit when assuming totally inefficient mass-transfer, it follows that to be able to explain such high Be fractions, mass-transfer must be non-conservative. 

We have highlighted the distinction between the initial binary fraction and the binary fraction that one is able to observe, and argued that these two quantities are not equal. This is so because a population of binary stars will always contain post-interaction systems that will appear to be single stars. The calculations outlined in this work provide rough constraints on this discrepancy, suggesting that the initial binary fraction is much higher than previously thought.


In conclusion, our theoretical argument serves to reinforce numerous observational arguments that suggest binary interactions to be responsible for Be stars. We conclude that observations of Be stars in young open clusters \citep{MiloneObs,2020A&A...634A..51B} do not contradict the hypothesis that Be stars originate exclusively from mass-transfer in binary systems. We have shown that if all Be stars are binary interaction products, somewhat extreme assumptions must be realised such as an initial binary fraction very close to unity, a shallow initial mass function and very non-conservative mass transfer. Whether or not these conditions can be met by the stars in the sky remains to be seen.

\bigbreak 
Acknowledgements: {The authors extend gratitude to an anonymous referee for useful comments on an earlier version of this manuscript.}

\bibliography{bib}

\begin{thebibliography}{101}
\expandafter\ifx\csname natexlab\endcsname\relax\def\natexlab#1{#1}\fi

\bibitem[{{Ahmed} \& {Sigut}(2017)}]{2017MNRAS.471.3398A}
{Ahmed}, A. \& {Sigut}, T.~A.~A. 2017,
  \href{http://dx.doi.org/10.1093/mnras/stx1737}{\color{magenta}\mnras},
  \href{https://ui.adsabs.harvard.edu/abs/2017MNRAS.471.3398A}{471, 3398}

\bibitem[{{Bestenlehner} {et~al.}(2020){Bestenlehner}, {Crowther},
  {Caballero-Nieves}, {Schneider}, {Sim{\'o}n-D{\'\i}az}, {Brands}, {de Koter},
  {Gr{\"a}fener}, {Herrero}, {Langer}, {Lennon}, {Ma{\'\i}z Apell{\'a}niz},
  {Puls}, \& {Vink}}]{2020MNRAS.tmp.2627B}
{Bestenlehner}, J.~M., {Crowther}, P.~A., {Caballero-Nieves}, S.~M., {et~al.}
  2020, \href{http://dx.doi.org/10.1093/mnras/staa2801}{\color{magenta}\mnras}
  \href{https://ui.adsabs.harvard.edu/abs/2020MNRAS.tmp.2627B}{[\eprint[arXiv]{2009.05136}]}

\bibitem[{{Bodensteiner} {et~al.}(2020{\natexlab{a}}){Bodensteiner}, {Sana},
  {Mahy}, {Patrick}, {de Koter}, {de Mink}, {Evans}, {G{\"o}tberg}, {Langer},
  {Lennon}, {Schneider}, \& {Tramper}}]{2020A&A...634A..51B}
{Bodensteiner}, J., {Sana}, H., {Mahy}, L., {et~al.} 2020{\natexlab{a}},
  \href{http://dx.doi.org/10.1051/0004-6361/201936743}{\color{magenta}\aap},
  \href{https://ui.adsabs.harvard.edu/abs/2020A&A...634A..51B}{634, A51}

\bibitem[{{Bodensteiner} {et~al.}(2020{\natexlab{b}}){Bodensteiner}, {Shenar},
  \& {Sana}}]{2020arXiv200613229B}
{Bodensteiner}, J., {Shenar}, T., \& {Sana}, H. 2020{\natexlab{b}},
  \href{http://dx.doi.org/10.1051/0004-6361/202037640}{\color{magenta}\aap},
  \href{https://ui.adsabs.harvard.edu/abs/2020A&A...641A..42B}{641, A42}

\bibitem[{{Boubert} \& {Evans}(2018)}]{2018MNRAS.477.5261B}
{Boubert}, D. \& {Evans}, N.~W. 2018,
  \href{http://dx.doi.org/10.1093/mnras/sty980}{\color{magenta}\mnras},
  \href{https://ui.adsabs.harvard.edu/abs/2018MNRAS.477.5261B}{477, 5261}

\bibitem[{{Brandt} \& {Podsiadlowski}(1995)}]{1995MNRAS.274..461B}
{Brandt}, N. \& {Podsiadlowski}, P. 1995,
  \href{http://dx.doi.org/10.1093/mnras/274.2.461}{\color{magenta}\mnras},
  \href{https://ui.adsabs.harvard.edu/abs/1995MNRAS.274..461B}{274, 461}

\bibitem[{{Brott} {et~al.}(2011){Brott}, {de Mink}, {Cantiello}, {Langer}, {de
  Koter}, {Evans}, {Hunter}, {Trundle}, \& {Vink}}]{2011A&A...530A.115B}
{Brott}, I., {de Mink}, S.~E., {Cantiello}, M., {et~al.} 2011,
  \href{http://dx.doi.org/10.1051/0004-6361/201016113}{\color{magenta}\aap},
  \href{https://ui.adsabs.harvard.edu/abs/2011A&A...530A.115B}{530, A115}

\bibitem[{{Bro{\v{z}}} {et~al.}(2021){Bro{\v{z}}}, {Mourard}, {Budaj},
  {Harmanec}, {Schmitt}, {Tallon-Bosc}, {Bonneau}, {Bo{\v{z}}i{\'c}}, {Gies},
  \& {{\v{S}}lechta}}]{2021A&A...645A..51B}
{Bro{\v{z}}}, M., {Mourard}, D., {Budaj}, J., {et~al.} 2021,
  \href{http://dx.doi.org/10.1051/0004-6361/202039035}{\color{magenta}\aap},
  \href{https://ui.adsabs.harvard.edu/abs/2021A&A...645A..51B}{645, A51}

\bibitem[{{Carciofi} {et~al.}(2012){Carciofi}, {Bjorkman}, {Otero}, {Okazaki},
  {{\v{S}}tefl}, {Rivinius}, {Baade}, \& {Haubois}}]{2012ApJ...744L..15C}
{Carciofi}, A.~C., {Bjorkman}, J.~E., {Otero}, S.~A., {et~al.} 2012,
  \href{http://dx.doi.org/10.1088/2041-8205/744/1/L15}{\color{magenta}\apjl},
  \href{https://ui.adsabs.harvard.edu/abs/2012ApJ...744L..15C}{744, L15}

\bibitem[{{Castro} {et~al.}(2014){Castro}, {Fossati}, {Langer},
  {Sim{\'o}n-D{\'\i}az}, {Schneider}, \& {Izzard}}]{2014A&A...570L..13C}
{Castro}, N., {Fossati}, L., {Langer}, N., {et~al.} 2014,
  \href{http://dx.doi.org/10.1051/0004-6361/201425028}{\color{magenta}\aap},
  \href{https://ui.adsabs.harvard.edu/abs/2014A&A...570L..13C}{570, L13}

\bibitem[{{Choi} {et~al.}(2016){Choi}, {Dotter}, {Conroy}, {Cantiello},
  {Paxton}, \& {Johnson}}]{2016ApJ...823..102C}
{Choi}, J., {Dotter}, A., {Conroy}, C., {et~al.} 2016,
  \href{http://dx.doi.org/10.3847/0004-637X/823/2/102}{\color{magenta}\apj},
  \href{https://ui.adsabs.harvard.edu/abs/2016ApJ...823..102C}{823, 102}

\bibitem[{{Claret} \& {Torres}(2016)}]{2016A&A...592A..15C}
{Claret}, A. \& {Torres}, G. 2016,
  \href{http://dx.doi.org/10.1051/0004-6361/201628779}{\color{magenta}\aap},
  \href{https://ui.adsabs.harvard.edu/abs/2016A&A...592A..15C}{592, A15}

\bibitem[{{Coe} {et~al.}(2020){Coe}, {Kennea}, {Evans}, \&
  {Udalski}}]{2020MNRAS.497L..50C}
{Coe}, M.~J., {Kennea}, J.~A., {Evans}, P.~A., \& {Udalski}, A. 2020,
  \href{http://dx.doi.org/10.1093/mnrasl/slaa112}{\color{magenta}\mnras},
  \href{https://ui.adsabs.harvard.edu/abs/2020MNRAS.497L..50C}{497, L50}

\bibitem[{{Collins} \& {Truax}(1995)}]{1995ApJ...439..860C}
{Collins}, George~W., I. \& {Truax}, R.~J. 1995,
  \href{http://dx.doi.org/10.1086/175225}{\color{magenta}\apj},
  \href{https://ui.adsabs.harvard.edu/abs/1995ApJ...439..860C}{439, 860}

\bibitem[{{de Mink} {et~al.}(2011{\natexlab{a}}){de Mink}, {Langer}, \&
  {Izzard}}]{2011BSRSL..80..543D}
{de Mink}, S.~E., {Langer}, N., \& {Izzard}, R.~G. 2011{\natexlab{a}}, Bulletin
  de la Societe Royale des Sciences de Liege,
  \href{https://ui.adsabs.harvard.edu/abs/2011BSRSL..80..543D}{80, 543}

\bibitem[{{de Mink} {et~al.}(2011{\natexlab{b}}){de Mink}, {Langer}, \&
  {Izzard}}]{2011IAUS..272..531D}
{de Mink}, S.~E., {Langer}, N., \& {Izzard}, R.~G. 2011{\natexlab{b}}, in
  Active OB Stars: Structure, Evolution, Mass Loss, and Critical Limits, ed.
  C.~{Neiner}, G.~{Wade}, G.~{Meynet}, \& G.~{Peters}, Vol. 272,
  \href{https://ui.adsabs.harvard.edu/abs/2011IAUS..272..531D}{531--532}

\bibitem[{{de Mink} {et~al.}(2007){de Mink}, {Pols}, \&
  {Hilditch}}]{2007A&A...467.1181D}
{de Mink}, S.~E., {Pols}, O.~R., \& {Hilditch}, R.~W. 2007,
  \href{http://dx.doi.org/10.1051/0004-6361:20067007}{\color{magenta}\aap},
  \href{https://ui.adsabs.harvard.edu/abs/2007A&A...467.1181D}{467, 1181}

\bibitem[{{de Mink} {et~al.}(2014){de Mink}, {Sana}, {Langer}, {Izzard}, \&
  {Schneider}}]{2014ApJ...782....7D}
{de Mink}, S.~E., {Sana}, H., {Langer}, N., {Izzard}, R.~G., \& {Schneider},
  F.~R.~N. 2014,
  \href{http://dx.doi.org/10.1088/0004-637X/782/1/7}{\color{magenta}\apj},
  \href{https://ui.adsabs.harvard.edu/abs/2014ApJ...782....7D}{782, 7}

\bibitem[{{Dorigo Jones} {et~al.}(2020){Dorigo Jones}, {Oey}, {Paggeot},
  {Castro}, \& {Moe}}]{2020arXiv200903571D}
{Dorigo Jones}, J., {Oey}, M.~S., {Paggeot}, K., {Castro}, N., \& {Moe}, M.
  2020, \href{http://dx.doi.org/10.3847/1538-4357/abbc6b}{\color{magenta}\apj},
  \href{https://ui.adsabs.harvard.edu/abs/2020ApJ...903...43D}{903, 43}

\bibitem[{{Dufton} {et~al.}(2013){Dufton}, {Langer}, {Dunstall}, {Evans},
  {Brott}, {de Mink}, {Howarth}, {Kennedy}, {McEvoy}, {Potter},
  {Ram{\'\i}rez-Agudelo}, {Sana}, {Sim{\'o}n-D{\'\i}az}, {Taylor}, \&
  {Vink}}]{2013A&A...550A.109D}
{Dufton}, P.~L., {Langer}, N., {Dunstall}, P.~R., {et~al.} 2013,
  \href{http://dx.doi.org/10.1051/0004-6361/201220273}{\color{magenta}\aap},
  \href{https://ui.adsabs.harvard.edu/abs/2013A&A...550A.109D}{550, A109}

\bibitem[{{Dunstall} {et~al.}(2011){Dunstall}, {Brott}, {Dufton}, {Lennon},
  {Evans}, {Smartt}, \& {Hunter}}]{2011A&A...536A..65D}
{Dunstall}, P.~R., {Brott}, I., {Dufton}, P.~L., {et~al.} 2011,
  \href{http://dx.doi.org/10.1051/0004-6361/201117588}{\color{magenta}\aap},
  \href{https://ui.adsabs.harvard.edu/abs/2011A&A...536A..65D}{536, A65}

\bibitem[{{Ekstr{\"o}m} {et~al.}(2008){Ekstr{\"o}m}, {Meynet}, {Maeder}, \&
  {Barblan}}]{2008A&A...478..467E}
{Ekstr{\"o}m}, S., {Meynet}, G., {Maeder}, A., \& {Barblan}, F. 2008,
  \href{http://dx.doi.org/10.1051/0004-6361:20078095}{\color{magenta}\aap},
  \href{https://ui.adsabs.harvard.edu/abs/2008A&A...478..467E}{478, 467}

\bibitem[{{Espinosa Lara} \& {Rieutord}(2011)}]{2011A&A...533A..43E}
{Espinosa Lara}, F. \& {Rieutord}, M. 2011,
  \href{http://dx.doi.org/10.1051/0004-6361/201117252}{\color{magenta}\aap},
  \href{https://ui.adsabs.harvard.edu/abs/2011A&A...533A..43E}{533, A43}

\bibitem[{{Ferrario} {et~al.}(2009){Ferrario}, {Pringle}, {Tout}, \&
  {Wickramasinghe}}]{2009MNRAS.400L..71F}
{Ferrario}, L., {Pringle}, J.~E., {Tout}, C.~A., \& {Wickramasinghe}, D.~T.
  2009,
  \href{http://dx.doi.org/10.1111/j.1745-3933.2009.00765.x}{\color{magenta}\mnras},
  \href{https://ui.adsabs.harvard.edu/abs/2009MNRAS.400L..71F}{400, L71}

\bibitem[{{Glebbeek} \& {Pols}(2008)}]{2008A&A...488.1017G}
{Glebbeek}, E. \& {Pols}, O.~R. 2008,
  \href{http://dx.doi.org/10.1051/0004-6361:200809931}{\color{magenta}\aap},
  \href{https://ui.adsabs.harvard.edu/abs/2008A&A...488.1017G}{488, 1017}

\bibitem[{{Gossage} {et~al.}(2019){Gossage}, {Conroy}, {Dotter},
  {Cabrera-Ziri}, {Dolphin}, {Bastian}, {Dalcanton}, {Goudfrooij}, {Johnson},
  {Williams}, {Rosenfield}, {Kalirai}, \& {Fouesneau}}]{2019ApJ...887..199G}
{Gossage}, S., {Conroy}, C., {Dotter}, A., {et~al.} 2019,
  \href{http://dx.doi.org/10.3847/1538-4357/ab5717}{\color{magenta}\apj},
  \href{https://ui.adsabs.harvard.edu/abs/2019ApJ...887..199G}{887, 199}

\bibitem[{{Hastings} {et~al.}(2020){Hastings}, {Wang}, \&
  {Langer}}]{2020A&A...633A.165H}
{Hastings}, B., {Wang}, C., \& {Langer}, N. 2020,
  \href{http://dx.doi.org/10.1051/0004-6361/201937018}{\color{magenta}\aap},
  \href{https://ui.adsabs.harvard.edu/abs/2020A&A...633A.165H}{633, A165}

\bibitem[{{Hjellming} \& {Webbink}(1987)}]{1987ApJ...318..794H}
{Hjellming}, M.~S. \& {Webbink}, R.~F. 1987,
  \href{http://dx.doi.org/10.1086/165412}{\color{magenta}\apj},
  \href{https://ui.adsabs.harvard.edu/abs/1987ApJ...318..794H}{318, 794}

\bibitem[{{Hogeveen}(1991)}]{1991PhDT.......257H}
{Hogeveen}, S.~J. 1991,
  \href{https://ui.adsabs.harvard.edu/abs/1991PhDT.......257H}{{The mass-ratio
  distribution of binary stars}}, PhD thesis, -

\bibitem[{{Huang} {et~al.}(2010){Huang}, {Gies}, \& {McSwain}}]{HuangGiesObs}
{Huang}, W., {Gies}, D.~R., \& {McSwain}, M.~V. 2010,
  \href{http://dx.doi.org/10.1088/0004-637X/722/1/605}{\color{magenta}\apj},
  \href{http://adsabs.harvard.edu/abs/2010ApJ...722..605H}{722, 605}

\bibitem[{{Hurley} {et~al.}(2002){Hurley}, {Tout}, \&
  {Pols}}]{2002MNRAS.329..897H}
{Hurley}, J.~R., {Tout}, C.~A., \& {Pols}, O.~R. 2002,
  \href{http://dx.doi.org/10.1046/j.1365-8711.2002.05038.x}{\color{magenta}\mnras},
  \href{https://ui.adsabs.harvard.edu/abs/2002MNRAS.329..897H}{329, 897}

\bibitem[{{Iqbal} \& {Keller}(2013)}]{IqbalObs}
{Iqbal}, S. \& {Keller}, S.~C. 2013,
  \href{http://dx.doi.org/10.1093/mnras/stt1502}{\color{magenta}\mnras},
  \href{http://esoads.eso.org/abs/2013MNRAS.435.3103I}{435, 3103}

\bibitem[{{Justham} {et~al.}(2014){Justham}, {Podsiadlowski}, \&
  {Vink}}]{2014ApJ...796..121J}
{Justham}, S., {Podsiadlowski}, P., \& {Vink}, J.~S. 2014,
  \href{http://dx.doi.org/10.1088/0004-637X/796/2/121}{\color{magenta}\apj},
  \href{https://ui.adsabs.harvard.edu/abs/2014ApJ...796..121J}{796, 121}

\bibitem[{{Kamann} {et~al.}(2020){Kamann}, {Bastian}, {Gossage}, {Baade},
  {Cabrera-Ziri}, {Da Costa}, {de Mink}, {Georgy}, {Giesers}, {G{\"o}ttgens},
  {Hilker}, {Husser}, {Lardo}, {Larsen}, {Mackey}, {Martocchia}, {Mucciarelli},
  {Platais}, {Roth}, {Salaris}, {Usher}, \& {Yong}}]{2020MNRAS.492.2177K}
{Kamann}, S., {Bastian}, N., {Gossage}, S., {et~al.} 2020,
  \href{http://dx.doi.org/10.1093/mnras/stz3583}{\color{magenta}\mnras},
  \href{https://ui.adsabs.harvard.edu/abs/2020MNRAS.492.2177K}{492, 2177}

\bibitem[{{Keller} {et~al.}(1999){Keller}, {Wood}, \&
  {Bessell}}]{1999A&AS..134..489K}
{Keller}, S.~C., {Wood}, P.~R., \& {Bessell}, M.~S. 1999,
  \href{http://dx.doi.org/10.1051/aas:1999151}{\color{magenta}\aaps},
  \href{https://ui.adsabs.harvard.edu/abs/1999A&AS..134..489K}{134, 489}

\bibitem[{{Kippenhahn}(1974)}]{1974IAUS...66...20K}
{Kippenhahn}, R. 1974, in Late Stages of Stellar Evolution, ed. R.~J. {Tayler}
  \& J.~E. {Hesser}, Vol.~66,
  \href{https://ui.adsabs.harvard.edu/abs/1974IAUS...66...20K}{20}

\bibitem[{{Klement} {et~al.}(2019){Klement}, {Carciofi}, {Rivinius}, {Ignace},
  {Matthews}, {Torstensson}, {Gies}, {Vieira}, {Richardson}, {Domiciano de
  Souza}, {Bjorkman}, {Hallinan}, {Faes}, {Mota}, {Gullingsrud}, {de Breuck},
  {Kervella}, {Cur{\'e}}, \& {Gunawan}}]{2019ApJ...885..147K}
{Klement}, R., {Carciofi}, A.~C., {Rivinius}, T., {et~al.} 2019,
  \href{http://dx.doi.org/10.3847/1538-4357/ab48e7}{\color{magenta}\apj},
  \href{https://ui.adsabs.harvard.edu/abs/2019ApJ...885..147K}{885, 147}

\bibitem[{{Klement} {et~al.}(2017){Klement}, {Carciofi}, {Rivinius},
  {Matthews}, {Vieira}, {Ignace}, {Bjorkman}, {Mota}, {Faes}, {Bratcher},
  {Cur{\'e}}, \& {{\v{S}}tefl}}]{2017A&A...601A..74K}
{Klement}, R., {Carciofi}, A.~C., {Rivinius}, T., {et~al.} 2017,
  \href{http://dx.doi.org/10.1051/0004-6361/201629932}{\color{magenta}\aap},
  \href{https://ui.adsabs.harvard.edu/abs/2017A&A...601A..74K}{601, A74}

\bibitem[{{Klencki} {et~al.}(2021){Klencki}, {Nelemans}, {Istrate}, \&
  {Chruslinska}}]{2020arXiv200611286K}
{Klencki}, J., {Nelemans}, G., {Istrate}, A.~G., \& {Chruslinska}, M. 2021,
  \href{http://dx.doi.org/10.1051/0004-6361/202038707}{\color{magenta}\aap},
  \href{https://ui.adsabs.harvard.edu/abs/2021A&A...645A..54K}{645, A54}

\bibitem[{{Kobulnicky} \& {Fryer}(2007)}]{2007ApJ...670..747K}
{Kobulnicky}, H.~A. \& {Fryer}, C.~L. 2007,
  \href{http://dx.doi.org/10.1086/522073}{\color{magenta}\apj},
  \href{https://ui.adsabs.harvard.edu/abs/2007ApJ...670..747K}{670, 747}

\bibitem[{{K{\"o}hler} {et~al.}(2006){K{\"o}hler}, {Petr-Gotzens},
  {McCaughrean}, {Bouvier}, {Duch{\^e}ne}, {Quirrenbach}, \&
  {Zinnecker}}]{2006A&A...458..461K}
{K{\"o}hler}, R., {Petr-Gotzens}, M.~G., {McCaughrean}, M.~J., {et~al.} 2006,
  \href{http://dx.doi.org/10.1051/0004-6361:20054561}{\color{magenta}\aap},
  \href{https://ui.adsabs.harvard.edu/abs/2006A&A...458..461K}{458, 461}

\bibitem[{{Kouwenhoven} {et~al.}(2009){Kouwenhoven}, {Brown}, {Goodwin},
  {Portegies Zwart}, \& {Kaper}}]{2009A&A...493..979K}
{Kouwenhoven}, M.~B.~N., {Brown}, A.~G.~A., {Goodwin}, S.~P., {Portegies
  Zwart}, S.~F., \& {Kaper}, L. 2009,
  \href{http://dx.doi.org/10.1051/0004-6361:200810234}{\color{magenta}\aap},
  \href{https://ui.adsabs.harvard.edu/abs/2009A&A...493..979K}{493, 979}

\bibitem[{{Kouwenhoven} {et~al.}(2007){Kouwenhoven}, {Brown}, {Portegies
  Zwart}, \& {Kaper}}]{2007A&A...474...77K}
{Kouwenhoven}, M.~B.~N., {Brown}, A.~G.~A., {Portegies Zwart}, S.~F., \&
  {Kaper}, L. 2007,
  \href{http://dx.doi.org/10.1051/0004-6361:20077719}{\color{magenta}\aap},
  \href{https://ui.adsabs.harvard.edu/abs/2007A&A...474...77K}{474, 77}

\bibitem[{{Kriz} \& {Harmanec}(1975)}]{1975BAICz..26...65K}
{Kriz}, S. \& {Harmanec}, P. 1975, Bulletin of the Astronomical Institutes of
  Czechoslovakia,
  \href{https://ui.adsabs.harvard.edu/abs/1975BAICz..26...65K}{26, 65}

\bibitem[{{Labadie-Bartz} {et~al.}(2017){Labadie-Bartz}, {Pepper}, {McSwain},
  {Bjorkman}, {Bjorkman}, {Lund}, {Rodriguez}, {Stassun}, {Stevens}, {James},
  {Kuhn}, {Siverd}, \& {Beatty}}]{2017AJ....153..252L}
{Labadie-Bartz}, J., {Pepper}, J., {McSwain}, M.~V., {et~al.} 2017,
  \href{http://dx.doi.org/10.3847/1538-3881/aa6396}{\color{magenta}\aj},
  \href{https://ui.adsabs.harvard.edu/abs/2017AJ....153..252L}{153, 252}

\bibitem[{{Langer}(2012)}]{2012ARA&A..50..107L}
{Langer}, N. 2012,
  \href{http://dx.doi.org/10.1146/annurev-astro-081811-125534}{\color{magenta}\araa},
  \href{https://ui.adsabs.harvard.edu/abs/2012ARA&A..50..107L}{50, 107}

\bibitem[{{Langer} {et~al.}(2020){Langer}, {Sch{\"u}rmann}, {Stoll},
  {Marchant}, {Lennon}, {Mahy}, {de Mink}, {Quast}, {Riedel}, {Sana},
  {Schneider}, {Schootemeijer}, {Wang}, {Almeida}, {Bestenlehner},
  {Bodensteiner}, {Castro}, {Clark}, {Crowther}, {Dufton}, {Evans}, {Fossati},
  {Gr{\"a}fener}, {Grassitelli}, {Grin}, {Hastings}, {Herrero}, {de Koter},
  {Menon}, {Patrick}, {Puls}, {Renzo}, {Sand er}, {Schneider}, {Sen}, {Shenar},
  {Sim{\'o}n-D{\'\i}as}, {Tauris}, {Tramper}, {Vink}, \&
  {Xu}}]{2020A&A...638A..39L}
{Langer}, N., {Sch{\"u}rmann}, C., {Stoll}, K., {et~al.} 2020,
  \href{http://dx.doi.org/10.1051/0004-6361/201937375}{\color{magenta}\aap},
  \href{https://ui.adsabs.harvard.edu/abs/2020A&A...638A..39L}{638, A39}

\bibitem[{{Lennon} {et~al.}(2005){Lennon}, {Lee}, {Dufton}, \&
  {Ryans}}]{2005A&A...438..265L}
{Lennon}, D.~J., {Lee}, J.~K., {Dufton}, P.~L., \& {Ryans}, R.~S.~I. 2005,
  \href{http://dx.doi.org/10.1051/0004-6361:20041653}{\color{magenta}\aap},
  \href{https://ui.adsabs.harvard.edu/abs/2005A&A...438..265L}{438, 265}

\bibitem[{{Li} {et~al.}(2012){Li}, {Kong}, {Charles}, {Lu}, {Bartlett}, {Coe},
  {McBride}, {Rajoelimanana}, {Udalski}, {Masetti}, \&
  {Franzen}}]{2012ApJ...761...99L}
{Li}, K.~L., {Kong}, A. K.~H., {Charles}, P.~A., {et~al.} 2012,
  \href{http://dx.doi.org/10.1088/0004-637X/761/2/99}{\color{magenta}\apj},
  \href{https://ui.adsabs.harvard.edu/abs/2012ApJ...761...99L}{761, 99}

\bibitem[{{Liu} {et~al.}(2006){Liu}, {van Paradijs}, \& {van den
  Heuvel}}]{2006A&A...455.1165L}
{Liu}, Q.~Z., {van Paradijs}, J., \& {van den Heuvel}, E.~P.~J. 2006,
  \href{http://dx.doi.org/10.1051/0004-6361:20064987}{\color{magenta}\aap},
  \href{https://ui.adsabs.harvard.edu/abs/2006A&A...455.1165L}{455, 1165}

\bibitem[{{Maeder} {et~al.}(1999){Maeder}, {Grebel}, \&
  {Mermilliod}}]{MaederObs}
{Maeder}, A., {Grebel}, E.~K., \& {Mermilliod}, J.-C. 1999, \aap,
  \href{http://adsabs.harvard.edu/abs/1999A%26A...346..459M}{346, 459}

\bibitem[{{Marino} {et~al.}(2018){Marino}, {Przybilla}, {Milone}, {Da Costa},
  {D'Antona}, {Dotter}, \& {Dupree}}]{2018AJ....156..116M}
{Marino}, A.~F., {Przybilla}, N., {Milone}, A.~P., {et~al.} 2018,
  \href{http://dx.doi.org/10.3847/1538-3881/aad3cd}{\color{magenta}\aj},
  \href{https://ui.adsabs.harvard.edu/abs/2018AJ....156..116M}{156, 116}

\bibitem[{{Martayan} {et~al.}(2010){Martayan}, {Baade}, \&
  {Fabregat}}]{MatayanObs}
{Martayan}, C., {Baade}, D., \& {Fabregat}, J. 2010,
  \href{http://dx.doi.org/10.1051/0004-6361/200911672}{\color{magenta}\aap},
  \href{http://adsabs.harvard.edu/abs/2010A%26A...509A..11M}{509, A11}

\bibitem[{{McSwain} \& {Gies}(2005)}]{2005ApJS..161..118M}
{McSwain}, M.~V. \& {Gies}, D.~R. 2005,
  \href{http://dx.doi.org/10.1086/432757}{\color{magenta}\apjs},
  \href{https://ui.adsabs.harvard.edu/abs/2005ApJS..161..118M}{161, 118}

\bibitem[{{Milone} {et~al.}(2018){Milone}, {Marino}, {Di Criscienzo},
  {D'Antona}, {Bedin}, {Da Costa}, {Piotto}, {Tailo}, {Dotter}, {Angeloni},
  {Anderson}, {Jerjen}, {Li}, {Dupree}, {Granata}, {Lagioia}, {Mackey},
  {Nardiello}, \& {Vesperini}}]{MiloneObs}
{Milone}, A.~P., {Marino}, A.~F., {Di Criscienzo}, M., {et~al.} 2018,
  \href{http://dx.doi.org/10.1093/mnras/sty661}{\color{magenta}\mnras},
  \href{https://ui.adsabs.harvard.edu/abs/2018MNRAS.477.2640M}{477, 2640}

\bibitem[{{Packet}(1981)}]{1981A&A...102...17P}
{Packet}, W. 1981, \aap,
  \href{https://ui.adsabs.harvard.edu/abs/1981A%26A...102...17P}{102, 17}

\bibitem[{{Pavlovskii} \& {Ivanova}(2015)}]{2015MNRAS.449.4415P}
{Pavlovskii}, K. \& {Ivanova}, N. 2015,
  \href{http://dx.doi.org/10.1093/mnras/stv619}{\color{magenta}\mnras},
  \href{https://ui.adsabs.harvard.edu/abs/2015MNRAS.449.4415P}{449, 4415}

\bibitem[{{Paxton} {et~al.}(2011){Paxton}, {Bildsten}, {Dotter}, {Herwig},
  {Lesaffre}, \& {Timmes}}]{Paxton2011}
{Paxton}, B., {Bildsten}, L., {Dotter}, A., {et~al.} 2011,
  \href{http://dx.doi.org/10.1088/0067-0049/192/1/3}{\color{magenta}\apjs},
  \href{https://ui.adsabs.harvard.edu/abs/2011ApJS..192....3P}{192, 3}

\bibitem[{{Paxton} {et~al.}(2013){Paxton}, {Cantiello}, {Arras}, {Bildsten},
  {Brown}, {Dotter}, {Mankovich}, {Montgomery}, {Stello}, {Timmes}, \&
  {Townsend}}]{Paxton2013}
{Paxton}, B., {Cantiello}, M., {Arras}, P., {et~al.} 2013,
  \href{http://dx.doi.org/10.1088/0067-0049/208/1/4}{\color{magenta}\apjs},
  \href{https://ui.adsabs.harvard.edu/abs/2013ApJS..208....4P}{208, 4}

\bibitem[{{Paxton} {et~al.}(2015){Paxton}, {Marchant}, {Schwab}, {Bauer},
  {Bildsten}, {Cantiello}, {Dessart}, {Farmer}, {Hu}, {Langer}, {Townsend},
  {Townsley}, \& {Timmes}}]{Paxton2015}
{Paxton}, B., {Marchant}, P., {Schwab}, J., {et~al.} 2015,
  \href{http://dx.doi.org/10.1088/0067-0049/220/1/15}{\color{magenta}\apjs},
  \href{https://ui.adsabs.harvard.edu/abs/2015ApJS..220...15P}{220, 15}

\bibitem[{{Paxton} {et~al.}(2018){Paxton}, {Schwab}, {Bauer}, {Bildsten},
  {Blinnikov}, {Duffell}, {Farmer}, {Goldberg}, {Marchant}, {Sorokina},
  {Thoul}, {Townsend}, \& {Timmes}}]{Paxton2018}
{Paxton}, B., {Schwab}, J., {Bauer}, E.~B., {et~al.} 2018,
  \href{http://dx.doi.org/10.3847/1538-4365/aaa5a8}{\color{magenta}\apjs},
  \href{https://ui.adsabs.harvard.edu/abs/2018ApJS..234...34P}{234, 34}

\bibitem[{{Paxton} {et~al.}(2019){Paxton}, {Smolec}, {Schwab}, {Gautschy},
  {Bildsten}, {Cantiello}, {Dotter}, {Farmer}, {Goldberg}, {Jermyn}, {Kanbur},
  {Marchant}, {Thoul}, {Townsend}, {Wolf}, {Zhang}, \& {Timmes}}]{Paxton2019}
{Paxton}, B., {Smolec}, R., {Schwab}, J., {et~al.} 2019,
  \href{http://dx.doi.org/10.3847/1538-4365/ab2241}{\color{magenta}\apjs},
  \href{https://ui.adsabs.harvard.edu/abs/2019ApJS..243...10P}{243, 10}

\bibitem[{{Peters}(1989)}]{1989SSRv...50....9P}
{Peters}, G.~J. 1989,
  \href{http://dx.doi.org/10.1007/BF00215915}{\color{magenta}\ssr},
  \href{https://ui.adsabs.harvard.edu/abs/1989SSRv...50....9P}{50, 9}

\bibitem[{{Petrovic} {et~al.}(2005){Petrovic}, {Langer}, \& {van der
  Hucht}}]{2005A&A...435.1013P}
{Petrovic}, J., {Langer}, N., \& {van der Hucht}, K.~A. 2005,
  \href{http://dx.doi.org/10.1051/0004-6361:20042368}{\color{magenta}\aap},
  \href{https://ui.adsabs.harvard.edu/abs/2005A&A...435.1013P}{435, 1013}

\bibitem[{{Pinsonneault} {et~al.}(1989){Pinsonneault}, {Kawaler}, {Sofia}, \&
  {Demarque}}]{1989ApJ...338..424P}
{Pinsonneault}, M.~H., {Kawaler}, S.~D., {Sofia}, S., \& {Demarque}, P. 1989,
  \href{http://dx.doi.org/10.1086/167210}{\color{magenta}\apj},
  \href{https://ui.adsabs.harvard.edu/abs/1989ApJ...338..424P}{338, 424}

\bibitem[{{Pols} {et~al.}(1991){Pols}, {Cote}, {Waters}, \&
  {Heise}}]{PolsBinaryModels}
{Pols}, O.~R., {Cote}, J., {Waters}, L.~B.~F.~M., \& {Heise}, J. 1991, \aap,
  \href{http://adsabs.harvard.edu/abs/1991A%26A...241..419P}{241, 419}

\bibitem[{{Porter}(1996)}]{1996MNRAS.280L..31P}
{Porter}, J.~M. 1996,
  \href{http://dx.doi.org/10.1093/mnras/280.3.L31}{\color{magenta}\mnras},
  \href{https://ui.adsabs.harvard.edu/abs/1996MNRAS.280L..31P}{280, L31}

\bibitem[{{Porter} \& {Rivinius}(2003)}]{2003PASP..115.1153P}
{Porter}, J.~M. \& {Rivinius}, T. 2003,
  \href{http://dx.doi.org/10.1086/378307}{\color{magenta}\pasp},
  \href{https://ui.adsabs.harvard.edu/abs/2003PASP..115.1153P}{115, 1153}

\bibitem[{{Raguzova} \& {Popov}(2005)}]{BeXRBcat}
{Raguzova}, N.~V. \& {Popov}, S.~B. 2005,
  \href{http://dx.doi.org/10.1080/10556790500497311}{\color{magenta}Astronomical
  and Astrophysical Transactions},
  \href{http://adsabs.harvard.edu/abs/2005A%26AT...24..151R}{24, 151}

\bibitem[{{R{\'\i}mulo} {et~al.}(2018){R{\'\i}mulo}, {Carciofi}, {Vieira},
  {Rivinius}, {Faes}, {Figueiredo}, {Bjorkman}, {Georgy}, {Ghoreyshi}, \&
  {Soszy{\'n}ski}}]{2018MNRAS.476.3555R}
{R{\'\i}mulo}, L.~R., {Carciofi}, A.~C., {Vieira}, R.~G., {et~al.} 2018,
  \href{http://dx.doi.org/10.1093/mnras/sty431}{\color{magenta}\mnras},
  \href{https://ui.adsabs.harvard.edu/abs/2018MNRAS.476.3555R}{476, 3555}

\bibitem[{{Salpeter}(1955)}]{1955ApJ...121..161S}
{Salpeter}, E.~E. 1955,
  \href{http://dx.doi.org/10.1086/145971}{\color{magenta}\apj},
  \href{https://ui.adsabs.harvard.edu/abs/1955ApJ...121..161S}{121, 161}

\bibitem[{{Sana} {et~al.}(2013){Sana}, {de Koter}, {de Mink}, {Dunstall},
  {Evans}, {H{\'e}nault-Brunet}, {Ma{\'\i}z Apell{\'a}niz},
  {Ram{\'\i}rez-Agudelo}, {Taylor}, {Walborn}, {Clark}, {Crowther}, {Herrero},
  {Gieles}, {Langer}, {Lennon}, \& {Vink}}]{2013A&A...550A.107S}
{Sana}, H., {de Koter}, A., {de Mink}, S.~E., {et~al.} 2013,
  \href{http://dx.doi.org/10.1051/0004-6361/201219621}{\color{magenta}\aap},
  \href{https://ui.adsabs.harvard.edu/abs/2013A&A...550A.107S}{550, A107}

\bibitem[{{Sana} {et~al.}(2012){Sana}, {de Mink}, {de Koter}, {Langer},
  {Evans}, {Gieles}, {Gosset}, {Izzard}, {Le Bouquin}, \&
  {Schneider}}]{2012Sci...337..444S}
{Sana}, H., {de Mink}, S.~E., {de Koter}, A., {et~al.} 2012,
  \href{http://dx.doi.org/10.1126/science.1223344}{\color{magenta}Science},
  \href{https://ui.adsabs.harvard.edu/abs/2012Sci...337..444S}{337, 444}

\bibitem[{{Scalo}(1986)}]{1986FCPh...11....1S}
{Scalo}, J.~M. 1986, \fcp,
  \href{https://ui.adsabs.harvard.edu/abs/1986FCPh...11....1S}{11, 1}

\bibitem[{{Schneider} {et~al.}(2015){Schneider}, {Izzard}, {Langer}, \& {de
  Mink}}]{2015ApJ...805...20S}
{Schneider}, F.~R.~N., {Izzard}, R.~G., {Langer}, N., \& {de Mink}, S.~E. 2015,
  \href{http://dx.doi.org/10.1088/0004-637X/805/1/20}{\color{magenta}\apj},
  \href{https://ui.adsabs.harvard.edu/abs/2015ApJ...805...20S}{805, 20}

\bibitem[{{Schneider} {et~al.}(2019){Schneider}, {Ohlmann}, {Podsiadlowski},
  {R{\"o}pke}, {Balbus}, {Pakmor}, \& {Springel}}]{2019Natur.574..211S}
{Schneider}, F. R.~N., {Ohlmann}, S.~T., {Podsiadlowski}, P., {et~al.} 2019,
  \href{http://dx.doi.org/10.1038/s41586-019-1621-5}{\color{magenta}\nat},
  \href{https://ui.adsabs.harvard.edu/abs/2019Natur.574..211S}{574, 211}

\bibitem[{{Schneider} {et~al.}(2018){Schneider}, {Sana}, {Evans},
  {Bestenlehner}, {Castro}, {Fossati}, {Gr{\"a}fener}, {Langer},
  {Ram{\'\i}rez-Agudelo}, {Sab{\'\i}n-Sanjuli{\'a}n}, {Sim{\'o}n-D{\'\i}az},
  {Tramper}, {Crowther}, {de Koter}, {de Mink}, {Dufton}, {Garcia}, {Gieles},
  {H{\'e}nault-Brunet}, {Herrero}, {Izzard}, {Kalari}, {Lennon}, {Ma{\'\i}z
  Apell{\'a}niz}, {Markova}, {Najarro}, {Podsiadlowski}, {Puls}, {Taylor}, {van
  Loon}, {Vink}, \& {Norman}}]{2018Sci...359...69S}
{Schneider}, F.~R.~N., {Sana}, H., {Evans}, C.~J., {et~al.} 2018,
  \href{http://dx.doi.org/10.1126/science.aan0106}{\color{magenta}Science},
  \href{https://ui.adsabs.harvard.edu/abs/2018Sci...359...69S}{359, 69}

\bibitem[{{Schootemeijer} {et~al.}(2018){Schootemeijer}, {G{\"o}tberg}, {de
  Mink}, {Gies}, \& {Zapartas}}]{2018A&A...615A..30S}
{Schootemeijer}, A., {G{\"o}tberg}, Y., {de Mink}, S.~E., {Gies}, D., \&
  {Zapartas}, E. 2018,
  \href{http://dx.doi.org/10.1051/0004-6361/201731194}{\color{magenta}\aap},
  \href{https://ui.adsabs.harvard.edu/abs/2018A&A...615A..30S}{615, A30}

\bibitem[{{Schootemeijer} {et~al.}(2019){Schootemeijer}, {Langer}, {Grin}, \&
  {Wang}}]{2019A&A...625A.132S}
{Schootemeijer}, A., {Langer}, N., {Grin}, N.~J., \& {Wang}, C. 2019,
  \href{http://dx.doi.org/10.1051/0004-6361/201935046}{\color{magenta}\aap},
  \href{https://ui.adsabs.harvard.edu/abs/2019A&A...625A.132S}{625, A132}

\bibitem[{{Secchi}(1866)}]{BeDiscovery}
{Secchi}, A. 1866,
  \href{http://dx.doi.org/10.1002/asna.18670680405}{\color{magenta}Astronomische
  Nachrichten}, \href{http://adsabs.harvard.edu/abs/1866AN.....68...63S}{68,
  63}

\bibitem[{{Shao} \& {Li}(2014)}]{2014ApJ...796...37S}
{Shao}, Y. \& {Li}, X.-D. 2014,
  \href{http://dx.doi.org/10.1088/0004-637X/796/1/37}{\color{magenta}\apj},
  \href{https://ui.adsabs.harvard.edu/abs/2014ApJ...796...37S}{796, 37}

\bibitem[{{Shao} \& {Li}(2016)}]{2016ApJ...833..108S}
{Shao}, Y. \& {Li}, X.-D. 2016,
  \href{http://dx.doi.org/10.3847/1538-4357/833/1/108}{\color{magenta}\apj},
  \href{https://ui.adsabs.harvard.edu/abs/2016ApJ...833..108S}{833, 108}

\bibitem[{{Shenar} {et~al.}(2020){Shenar}, {Bodensteiner}, {Abdul-Masih},
  {Fabry}, {Mahy}, {Marchant}, {Banyard}, {Bowman}, {Dsilva}, {Hawcroft},
  {Reggiani}, \& {Sana}}]{2020A&A...639L...6S}
{Shenar}, T., {Bodensteiner}, J., {Abdul-Masih}, M., {et~al.} 2020,
  \href{http://dx.doi.org/10.1051/0004-6361/202038275}{\color{magenta}\aap},
  \href{https://ui.adsabs.harvard.edu/abs/2020A&A...639L...6S}{639, L6}

\bibitem[{{Sigut} {et~al.}(2009){Sigut}, {McGill}, \&
  {Jones}}]{2009ApJ...699.1973S}
{Sigut}, T.~A.~A., {McGill}, M.~A., \& {Jones}, C.~E. 2009,
  \href{http://dx.doi.org/10.1088/0004-637X/699/2/1973}{\color{magenta}\apj},
  \href{https://ui.adsabs.harvard.edu/abs/2009ApJ...699.1973S}{699, 1973}

\bibitem[{{Sternberg} {et~al.}(2003){Sternberg}, {Hoffmann}, \&
  {Pauldrach}}]{2003ApJ...599.1333S}
{Sternberg}, A., {Hoffmann}, T.~L., \& {Pauldrach}, A.~W.~A. 2003,
  \href{http://dx.doi.org/10.1086/379506}{\color{magenta}\apj},
  \href{https://ui.adsabs.harvard.edu/abs/2003ApJ...599.1333S}{599, 1333}

\bibitem[{{Struve}(1931)}]{Struve}
{Struve}, O. 1931,
  \href{http://dx.doi.org/10.1086/143298}{\color{magenta}\apj},
  \href{http://adsabs.harvard.edu/abs/1931ApJ....73...94S}{73, 94}

\bibitem[{{Sun} {et~al.}(2019){Sun}, {de Grijs}, {Deng}, \&
  {Albrow}}]{2019ApJ...876..113S}
{Sun}, W., {de Grijs}, R., {Deng}, L., \& {Albrow}, M.~D. 2019,
  \href{http://dx.doi.org/10.3847/1538-4357/ab16e4}{\color{magenta}\apj},
  \href{https://ui.adsabs.harvard.edu/abs/2019ApJ...876..113S}{876, 113}

\bibitem[{{Tokovinin} \& {Moe}(2020)}]{2020MNRAS.491.5158T}
{Tokovinin}, A. \& {Moe}, M. 2020,
  \href{http://dx.doi.org/10.1093/mnras/stz3299}{\color{magenta}\mnras},
  \href{https://ui.adsabs.harvard.edu/abs/2020MNRAS.491.5158T}{491, 5158}

\bibitem[{{Tout}(1991)}]{1991MNRAS.250..701T}
{Tout}, C.~A. 1991,
  \href{http://dx.doi.org/10.1093/mnras/250.4.701}{\color{magenta}\mnras},
  \href{https://ui.adsabs.harvard.edu/abs/1991MNRAS.250..701T}{250, 701}

\bibitem[{{Tout} {et~al.}(1997){Tout}, {Aarseth}, {Pols}, \&
  {Eggleton}}]{1997MNRAS.291..732T}
{Tout}, C.~A., {Aarseth}, S.~J., {Pols}, O.~R., \& {Eggleton}, P.~P. 1997,
  \href{http://dx.doi.org/10.1093/mnras/291.4.732}{\color{magenta}\mnras},
  \href{https://ui.adsabs.harvard.edu/abs/1997MNRAS.291..732T}{291, 732}

\bibitem[{{Townsend} {et~al.}(2004){Townsend}, {Owocki}, \&
  {Howarth}}]{Townsend}
{Townsend}, R.~H.~D., {Owocki}, S.~P., \& {Howarth}, I.~D. 2004,
  \href{http://dx.doi.org/10.1111/j.1365-2966.2004.07627.x}{\color{magenta}\mnras},
  \href{http://adsabs.harvard.edu/abs/2004MNRAS.350..189T}{350, 189}

\bibitem[{{Trimble}(1990)}]{1990MNRAS.242...79T}
{Trimble}, V. 1990,
  \href{http://dx.doi.org/10.1093/mnras/242.1.79}{\color{magenta}\mnras},
  \href{https://ui.adsabs.harvard.edu/abs/1990MNRAS.242...79T}{242, 79}

\bibitem[{{van Bever} \& {Vanbeveren}(1997)}]{1997A&A...322..116V}
{van Bever}, J. \& {Vanbeveren}, D. 1997, \aap,
  \href{https://ui.adsabs.harvard.edu/abs/1997A&A...322..116V}{322, 116}

\bibitem[{{Vinciguerra} {et~al.}(2020){Vinciguerra}, {Neijssel},
  {Vigna-G{\'o}mez}, {Mandel}, {Podsiadlowski}, {Maccarone}, {Nicholl},
  {Kingdon}, {Perry}, \& {Salemi}}]{2020arXiv200300195V}
{Vinciguerra}, S., {Neijssel}, C.~J., {Vigna-G{\'o}mez}, A., {et~al.} 2020,
  \href{http://dx.doi.org/10.1093/mnras/staa2177}{\color{magenta}\mnras},
  \href{https://ui.adsabs.harvard.edu/abs/2020MNRAS.498.4705V}{498, 4705}

\bibitem[{{von Zeipel}(1924)}]{1924MNRAS..84..665V}
{von Zeipel}, H. 1924,
  \href{http://dx.doi.org/10.1093/mnras/84.9.665}{\color{magenta}\mnras},
  \href{http://adsabs.harvard.edu/abs/1924MNRAS..84..665V}{84, 665}

\bibitem[{{Wang} {et~al.}(2020){Wang}, {Langer}, {Schootemeijer}, {Castro},
  {Adscheid}, {Marchant}, \& {Hastings}}]{2020ApJ...888L..12W}
{Wang}, C., {Langer}, N., {Schootemeijer}, A., {et~al.} 2020,
  \href{http://dx.doi.org/10.3847/2041-8213/ab6171}{\color{magenta}\apjl},
  \href{https://ui.adsabs.harvard.edu/abs/2020ApJ...888L..12W}{888, L12}

\bibitem[{{Wellstein} {et~al.}(2001){Wellstein}, {Langer}, \&
  {Braun}}]{2001A&A...369..939W}
{Wellstein}, S., {Langer}, N., \& {Braun}, H. 2001,
  \href{http://dx.doi.org/10.1051/0004-6361:20010151}{\color{magenta}\aap},
  \href{https://ui.adsabs.harvard.edu/abs/2001A&A...369..939W}{369, 939}

\bibitem[{{Wickramasinghe} {et~al.}(2014){Wickramasinghe}, {Tout}, \&
  {Ferrario}}]{2014MNRAS.437..675W}
{Wickramasinghe}, D.~T., {Tout}, C.~A., \& {Ferrario}, L. 2014,
  \href{http://dx.doi.org/10.1093/mnras/stt1910}{\color{magenta}\mnras},
  \href{https://ui.adsabs.harvard.edu/abs/2014MNRAS.437..675W}{437, 675}

\bibitem[{{Zahn} {et~al.}(2010){Zahn}, {Ranc}, \&
  {Morel}}]{2010A&A...517A...7Z}
{Zahn}, J.~P., {Ranc}, C., \& {Morel}, P. 2010,
  \href{http://dx.doi.org/10.1051/0004-6361/200913817}{\color{magenta}\aap},
  \href{https://ui.adsabs.harvard.edu/abs/2010A&A...517A...7Z}{517, A7}

\bibitem[{{Zorec} {et~al.}(2005){Zorec}, {Fr{\'e}mat}, \&
  {Cidale}}]{2005A&A...441..235Z}
{Zorec}, J., {Fr{\'e}mat}, Y., \& {Cidale}, L. 2005,
  \href{http://dx.doi.org/10.1051/0004-6361:20053051}{\color{magenta}\aap},
  \href{https://ui.adsabs.harvard.edu/abs/2005A&A...441..235Z}{441, 235}

\bibitem[{{Zorec} {et~al.}(2016){Zorec}, {Fr{\'e}mat}, {Domiciano de Souza},
  {Royer}, {Cidale}, {Hubert}, {Semaan}, {Martayan}, {Cochetti}, {Arias},
  {Aidelman}, \& {Stee}}]{2016A&A...595A.132Z}
{Zorec}, J., {Fr{\'e}mat}, Y., {Domiciano de Souza}, A., {et~al.} 2016,
  \href{http://dx.doi.org/10.1051/0004-6361/201628760}{\color{magenta}\aap},
  \href{https://ui.adsabs.harvard.edu/abs/2016A&A...595A.132Z}{595, A132}

\end{thebibliography}
\bibliographystyle{aa_url}
\setlength{\bibsep}{0pt}

\clearpage


\begin{appendix}

\section{Stars at the critical velocity \label{App2}}
Here we investigate how critically rotating and slowly rotating stars vary in terms of their effective temperatures and luminosities.  Owing to the fact that stable and reliable numerical models of very fast rotating stars are difficult to produce, we shall employ a simple analytic model.

The luminosity of a main-sequence star is generated from nuclear reactions in the central region. As stars are centrally condensed, when a given star is spun up to the critical velocity the centripetal forces acting in the central regions are much weaker than the force of gravity, meaning the structure of the core is largely unchanged. Hence, the intrinsic luminosity is constant to first order.

What does however change when a star is spun up is the outer structure. The equatorial radius increases, and therefore so does the surface area of the star, $S$. This then causes the effective temperature to decrease, as evidenced by the Stefan–Boltzmann law
\begin{align}
L \propto S T_{\rm{eff}}^4 . \label{Eq:stefanboltzmann}
\end{align}

To characterise this change in effective temperature we use the Roche model, which describes a star with all mass concentrated at the centre, and rotating with constant angular velocity $\Omega$. In this framework, the effective potential, with respect to the radial coordinate $r$ and latitude $\theta$, is 
\begin{align}
\Psi(r, \theta) = -\frac{GM}{r} - \frac{1}{2} \Omega^2 r^2 sin^2(\theta).
\end{align}

At the critical angular velocity, the polar radius, $R_p$ is equal to the radius of an equivalent non-rotating star of the same mass, whereas the equatorial radius, $R_e$ is given by 
\begin{align}
R_e=\frac{3}{2} R_p.
\end{align}

Taking the $x-y$ plane to be parallel to the axis of rotation, where $y$ represents the distance along the rotation axis and $x$ the perpendicular distance from the rotation axis, the surface of a critically rotating star is described by
\begin{align}
\left(\frac{y}{R_e}\right)^2= \left(\frac{2}{3-x^2 /R_e^2}\right)^2 - \left(\frac{x}{R_e}\right)^2, \label{eq:s}
\end{align}
\citep{2010A&A...517A...7Z}
The surface area of a star is \citep[see][Appendix B]{Paxton2019} , in general
\begin{align}
S = 4 \pi \int _0 ^ {R_e} x \sqrt{\left(\frac{dy}{dx}\right)^2 +1} \, dx \label{eq:sa}.
\end{align}
After choosing units such that $R_e =1$, Eqs. \ref{eq:s} and \ref{eq:sa} can be solved numerically to give the surface area of a critical rotator, $S_c$, as
\begin{align}
S_c \approx 4 \pi \times 0.7028. 
\end{align}
Compare this to the surface areas of non-rotating stars with radii $R_e$ and $R_p$,
\begin{align}
S_0 (r=R_e) = 4 \pi
\end{align}
and 
\begin{align}
S_0 (r=R_p) = 4 \pi \left(\frac{2}{3}\right)^2  \approx 4 \pi \times 0.4444.
\end{align}
As expected we have 
\begin{align}
S_o (r=R_p) < S_c < S_o (r=R_e).
\end{align}
We see that the surface area of a critically rotating star is around 1.58 times larger than that of its non-rotating counterpart. 

Therefore from Eq. \ref{Eq:stefanboltzmann}, the temperature of a star after having been spun up to critical decreases by a factor of 0.89.


\section{Stellar isochrones \label{App:isochrones}}

As our model predictions give the Be fraction as a function of mass, to make an effective comparison with observations, we must extract masses from stars in the colour-magnitude diagram. To this end, we employ isochrones of single rotating stars to assign mass ranges to different areas of the colour magnitude diagram. 

We use the grid of \citet{2019A&A...625A.132S} that has been extended to masses between 2 and 20\Msol with slight changes to internal mixing - see below.
The code used was MESA \citep{Paxton2011, Paxton2013, Paxton2015, Paxton2018, Paxton2019}.
Models were computed at initial rotation rates between 0 and 80\% of critical velocity in steps of 10\%. As is standard in MESA the critical velocity is defined as 
\begin{align}
v_{\textrm{crit}} = \sqrt{\frac{GM}{R} (1-\Gamma)}, \label{Eq:critvel}
\end{align} 
where $\Gamma$ is the ratio of luminosity to Eddington luminosity, and is negligible for the models presented here. 
During early times, the models undergo a relaxation period, during which the critical velocity fraction can oscillate wildly. To circumvent this, we define the initial critical velocity fraction at the point when the model has burnt 3\% of its initial hydrogen content by mass.

The physics employed in the models is mostly identical to that of \citet{2011A&A...530A.115B}, except for the treatment of two mixing processes. Stepped convective overshooting is adopted that extends the convective zone by $\alpha_{\textrm{OV}}$ times the local pressure scale-height. A dependence of $\alpha_{\textrm{OV}}$ with mass accounts for observational trends \citep{2014A&A...570L..13C,2016A&A...592A..15C,2019A&A...625A.132S}, whereby $\alpha_{\textrm{OV}}$ increases linearly from 0.1 at 1.66\Msun \cite{2016A&A...592A..15C} to 0.3 at 20\Msun \cite{2011A&A...530A.115B}. Furthermore, time smoothing in rotational mixing is turned off to avoid unrealistically strong mixing.

\begin{figure}
	\includegraphics[width=1.0\linewidth]{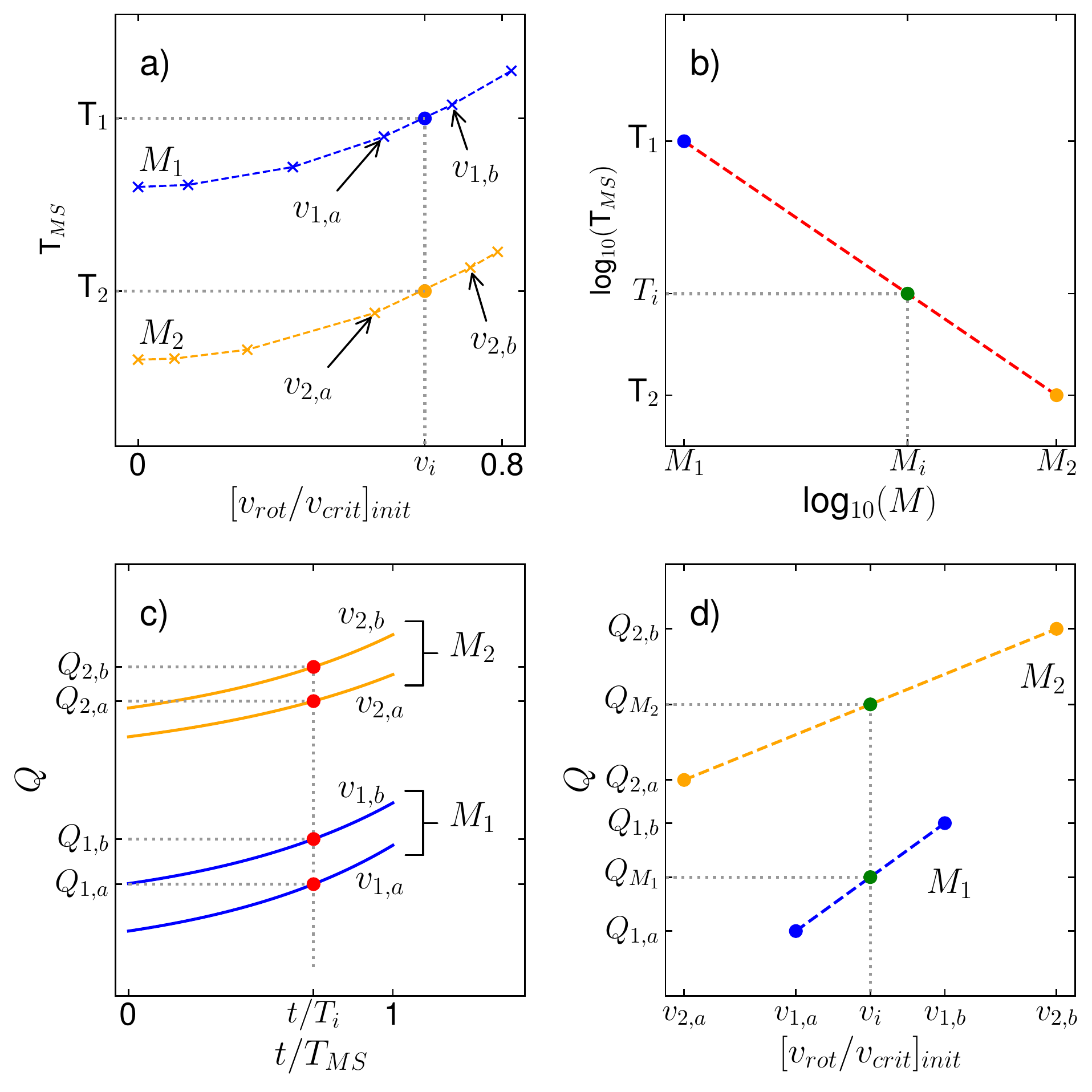}
	\centering
	\caption{Schematic representation of the interpolation procedure employed to produce isochrones. See text for a thorough explanation.}
	\label{fig:iso} 
\end{figure}

Isochrones are generated through a series of linear interpolations and are split up into two equivalent-evolutionary-phases (EEPs). The first phase lasts until core hydrogen depletion, and the second phase from core hydrogen depletion until core helium depletion. To compute the parameters of a star in the first EEP with initial mass $M_i$, initial critical velocity fraction $\varv_i$ at time $t$, we first find the time at which this star would experience core hydrogen exhaustion, $T_{MS}$. In total, four models are used for the interpolation, two models with initial masses $M_1$ and initial critical velocity fractions $\varv_{1,a}$ and $\varv_{1,b}$ and similarly two models with initial masses $M_2$ and initial critical velocity fractions $\varv_{2,a}$ and $\varv_{2,b}$. The models are selected such that $M_1 < M_i < M_2$ and $\varv_{1,a}< \varv_i < \varv_{1,b}$ and similarly for $\varv_{2,a}$, $\varv_{2,b}$. For $M_1$ and $M_2$ we interpolate the lifetime when initial \Vcritfrac = $\varv_i$ from these models as is shown in Fig. \ref{fig:iso} a. The hydrogen burning lifetime is then computed as an interpolation in mass between the values of $M_1$ and $M_2$, as depicted in Fig. \ref{fig:iso} b. For this step, the most accurate results are obtained when the logarithm of hydrogen burning lifetime is interpolated against the logarithm of initial mass. Using the interpolated lifetime, $T_i$, of this star with initial mass $M_i$ and initial critical velocity fraction $\varv_i$, we define its fractional lifetime as $t/T_i$. This fractional lifetime is the value at which all further interpolations will be carried out. Next, a given quantity (for the purposes of making isochrones, the quantities of interest are effective temperature and luminosity), $Q$, is interpolated at a fractional lifetime of $t/T_i$ for the four selected models, as in \ref{fig:iso} c. The penultimate step is to calculate the quantities $Q_{M_1}$, $Q_{M_2}$ which represent the values of $Q$ of a star with mass $M_1$, $M_2$, initial critical velocity fraction  $\varv_i$ and fractional lifetime $t/T_i$ by interpolating across initial critical velocity fraction like in Fig. \ref{fig:iso} d. Finally, an interpolation in initial mass between the quantities  $Q_{M_1}$ and $Q_{M_2}$ is done to produce the value of the chosen parameter for a star of given mass, initial rotation rate and age. 

To generate the second EEP, the same procedure is used but only with a different fractional lifetime, namely the fractional helium-burning lifetime, $t/T_{He}$ such that at core hydrogen exhaustion $t/T_{He}=0$ and at core helium exhaustion $t/T_{He}=1$.

Absolute magnitudes in Hubble Space Telescope filters are computed by interpolating tables of synthetic stellar spectra provided by the MIST project \citet{2016ApJ...823..102C}. Apparent magnitudes are then calculated  as 
\begin{align}
m_{\textrm{F814W}} = M_{\textrm{F814W}} +  A_{F814W} + \mu , \\
m_{\textrm{F336W}} = M_{\textrm{F336W}} +  A_{F336W} + \mu ,
\end{align}
with $\mu$ being the distance modulus and absorption coefficients $A_{F814W}=2.04 E(B-V)$ and $A_{F336W}=5.16 E(B-V)$ \citep{MiloneObs}.

\end{appendix}

\end{document}